\ifx\mnmacrosloaded\undefined 
%
%
%
%

\catcode `\@=11 

\def\@version{1.6}
\def\@verdate{18th September 1995}

%
%


\newif\ifprod@font

\ifx\@typeface\undefined
  \def\@typeface{Comp. Modern}\prod@fontfalse
\else
  \prod@fonttrue 
\fi

\def\newfam{\alloc@8\fam\chardef\sixt@@n} 

\ifprod@font
\font\fiverm=mtr10 at 5pt
\font\fivebf=mtbx10 at 5pt
\font\fiveit=mtti10 at 5pt
\font\fivesl=mtsl10 at 5pt
\font\fivett=cmtt8 at 5pt     \hyphenchar\fivett=-1
\font\fivecsc=mtcsc10 at 5pt
\font\fivesf=mtss10 at 5pt
\font\fivei=mtmi10 at 5pt      \skewchar\fivei='177
\font\fivesy=mtsy10 at 5pt     \skewchar\fivesy='60

\font\sixrm=mtr10 at 6pt
\font\sixbf=mtbx10 at 6pt
\font\sixit=mtti10 at 6pt
\font\sixsl=mtsl10 at 6pt
\font\sixtt=cmtt8 at 6pt      \hyphenchar\sixtt=-1
\font\sixcsc=mtcsc10 at 6pt
\font\sixsf=mtss10 at 6pt
\font\sixi=mtmi10 at 6pt       \skewchar\sixi='177
\font\sixsy=mtsy10 at 6pt      \skewchar\sixsy='60

\font\sevenrm=mtr10 at 7pt
\font\sevenbf=mtbx10 at 7pt
\font\sevenit=mtti10 at 7pt
\font\sevensl=mtsl10 at 7pt
\font\seventt=cmtt8 at 7pt     \hyphenchar\seventt=-1
\font\sevencsc=mtcsc10 at 7pt
\font\sevensf=mtss10 at 7pt
\font\seveni=mtmi10 at 7pt      \skewchar\seveni='177
\font\sevensy=mtsy10 at 7pt     \skewchar\sevensy='60

\font\eightrm=mtr10 at 8pt
\font\eightbf=mtbx10 at 8pt
\font\eightit=mtti10 at 8pt
\font\eighti=mtmi10 at 8pt      \skewchar\eighti='177
\font\eightsy=mtsy10 at 8pt     \skewchar\eightsy='60
\font\eightsl=mtsl10 at 8pt
\font\eighttt=cmtt8             \hyphenchar\eighttt=-1
\font\eightcsc=mtcsc10 at 8pt
\font\eightsf=mtss10 at 8pt

\font\ninerm=mtr10 at 9pt
\font\ninebf=mtbx10 at 9pt
\font\nineit=mtti10 at 9pt
\font\ninei=mtmi10 at 9pt      \skewchar\ninei='177
\font\ninesy=mtsy10 at 9pt     \skewchar\ninesy='60
\font\ninesl=mtsl10 at 9pt
\font\ninett=cmtt9             \hyphenchar\ninett=-1
\font\ninecsc=mtcsc10 at 9pt
\font\ninesf=mtss10 at 9pt

\font\tenrm=mtr10
\font\tenbf=mtbx10
\font\tenit=mtti10
\font\teni=mtmi10		\skewchar\teni='177
\font\tensy=mtsy10		\skewchar\tensy='60
\font\tenex=cmex10
\font\tensl=mtsl10
\font\tentt=cmtt10		\hyphenchar\tentt=-1
\font\tencsc=mtcsc10
\font\tensf=mtss10

\font\elevenrm=mtr10 at 11pt
\font\elevenbf=mtbx10 at 11pt
\font\elevenit=mtti10 at 11pt
\font\eleveni=mtmi10 at 11pt      \skewchar\eleveni='177
\font\elevensy=mtsy10 at 11pt     \skewchar\elevensy='60
\font\elevensl=mtsl10 at 11pt
\font\eleventt=cmtt10 at 11pt     \hyphenchar\eleventt=-1
\font\elevencsc=mtcsc10 at 11pt
\font\elevensf=mtss10 at 11pt

\font\twelverm=mtr10 at 12pt
\font\twelvebf=mtbx10 at 12pt
\font\twelveit=mtti10 at 12pt
\font\twelvesl=mtsl10 at 12pt
\font\twelvett=cmtt12             \hyphenchar\twelvett=-1
\font\twelvecsc=mtcsc10 at 12pt
\font\twelvesf=mtss10 at 12pt
\font\twelvei=mtmi10 at 12pt      \skewchar\twelvei='177
\font\twelvesy=mtsy10 at 12pt     \skewchar\twelvesy='60

\font\fourteenrm=mtr10 at 14pt
\font\fourteenbf=mtbx10 at 14pt
\font\fourteenit=mtti10 at 14pt
\font\fourteeni=mtmi10 at 14pt      \skewchar\fourteeni='177
\font\fourteensy=mtsy10 at 14pt     \skewchar\fourteensy='60
\font\fourteensl=mtsl10 at 14pt
\font\fourteentt=cmtt12 at 14pt     \hyphenchar\fourteentt=-1
\font\fourteencsc=mtcsc10 at 14pt
\font\fourteensf=mtss10 at 14pt

\font\seventeenrm=mtr10 at 17pt
\font\seventeenbf=mtbx10 at 17pt
\font\seventeenit=mtti10 at 17pt
\font\seventeeni=mtmi10 at 17pt      \skewchar\seventeeni='177
\font\seventeensy=mtsy10 at 17pt     \skewchar\seventeensy='60
\font\seventeensl=mtsl10 at 17pt
\font\seventeentt=cmtt12 at 17pt     \hyphenchar\seventeentt=-1
\font\seventeencsc=mtcsc10 at 17pt
\font\seventeensf=mtss10 at 17pt
\else
\font\fiverm=cmr5
\font\fivei=cmmi5             \skewchar\fivei='177
\font\fivesy=cmsy5            \skewchar\fivesy='60
\font\fivebf=cmbx5

\font\sixrm=cmr6
\font\sixi=cmmi6             \skewchar\sixi='177
\font\sixsy=cmsy6            \skewchar\sixsy='60
\font\sixbf=cmbx6

\font\sevenrm=cmr7
\font\sevenit=cmti7
\font\seveni=cmmi7             \skewchar\seveni='177
\font\sevensy=cmsy7            \skewchar\sevensy='60
\font\sevenbf=cmbx7

\font\eightrm=cmr8
\font\eightbf=cmbx8
\font\eightit=cmti8
\font\eighti=cmmi8			\skewchar\eighti='177
\font\eightsy=cmsy8			\skewchar\eightsy='60
\font\eightsl=cmsl8
\font\eighttt=cmtt8			\hyphenchar\eighttt=-1
\font\eightcsc=cmcsc10 at 8pt
\font\eightsf=cmss8

\font\ninerm=cmr9
\font\ninebf=cmbx9
\font\nineit=cmti9
\font\ninei=cmmi9			\skewchar\ninei='177
\font\ninesy=cmsy9			\skewchar\ninesy='60
\font\ninesl=cmsl9
\font\ninett=cmtt9			\hyphenchar\ninett=-1
\font\ninecsc=cmcsc10 at 9pt
\font\ninesf=cmss9

\font\tenrm=cmr10
\font\tenbf=cmbx10
\font\tenit=cmti10
\font\teni=cmmi10		\skewchar\teni='177
\font\tensy=cmsy10		\skewchar\tensy='60
\font\tenex=cmex10
\font\tensl=cmsl10
\font\tentt=cmtt10		\hyphenchar\tentt=-1
\font\tencsc=cmcsc10
\font\tensf=cmss10

\font\elevenrm=cmr10 scaled \magstephalf
\font\elevenbf=cmbx10 scaled \magstephalf
\font\elevenit=cmti10 scaled \magstephalf
\font\eleveni=cmmi10 scaled \magstephalf	\skewchar\eleveni='177
\font\elevensy=cmsy10 scaled \magstephalf	\skewchar\elevensy='60
\font\elevensl=cmsl10 scaled \magstephalf
\font\eleventt=cmtt10 scaled \magstephalf	\hyphenchar\eleventt=-1
\font\elevencsc=cmcsc10 scaled \magstephalf
\font\elevensf=cmss10 scaled \magstephalf

\font\twelverm=cmr10 scaled \magstep1
\font\twelvebf=cmbx10 scaled \magstep1
\font\twelvei=cmmi10 scaled \magstep1      \skewchar\twelvei='177
\font\twelvesy=cmsy10 scaled \magstep1     \skewchar\twelvesy='60

\font\fourteenrm=cmr10 scaled \magstep2
\font\fourteenbf=cmbx10 scaled \magstep2
\font\fourteenit=cmti10 scaled \magstep2
\font\fourteeni=cmmi10 scaled \magstep2		\skewchar\fourteeni='177
\font\fourteensy=cmsy10 scaled \magstep2	\skewchar\fourteensy='60
\font\fourteensl=cmsl10 scaled \magstep2
\font\fourteentt=cmtt10 scaled \magstep2	\hyphenchar\fourteentt=-1
\font\fourteencsc=cmcsc10 scaled \magstep2
\font\fourteensf=cmss10 scaled \magstep2

\font\seventeenrm=cmr10 scaled \magstep3
\font\seventeenbf=cmbx10 scaled \magstep3
\font\seventeenit=cmti10 scaled \magstep3
\font\seventeeni=cmmi10 scaled \magstep3	\skewchar\seventeeni='177
\font\seventeensy=cmsy10 scaled \magstep3	\skewchar\seventeensy='60
\font\seventeensl=cmsl10 scaled \magstep3
\font\seventeentt=cmtt10 scaled \magstep3	\hyphenchar\seventeentt=-1
\font\seventeencsc=cmcsc10 scaled \magstep3
\font\seventeensf=cmss10 scaled \magstep3
\fi

\def\hexnumber#1{\ifcase#1 0\or1\or2\or3\or4\or5\or6\or7\or8\or9\or
  A\or B\or C\or D\or E\or F\fi}

\def\makestrut{%
  \setbox\strutbox=\hbox{%
    \vrule height.7\baselineskip depth.3\baselineskip width \z@}%
}

\def\baselinestretch{1}
\newskip\tmp@bls

\def\b@ls#1{
  \tmp@bls=#1\relax
  \baselineskip=#1\relax\makestrut
  \normalbaselineskip=\baselinestretch\tmp@bls
  \normalbaselines
}

\def\nostb@ls#1{
  \normalbaselineskip=#1\relax
  \normalbaselines
  \makestrut
}

%

\newfam\scfam  
\newfam\sffam  

\def\mit{\fam\@ne}
\def\cal{\fam\tw@}
\def\em{\ifdim\fontdimen1\font>\z@ \rm\else\it\fi}

\textfont3=\tenex
\scriptfont3=\tenex
\scriptscriptfont3=\tenex

\setbox0=\hbox{\tenex B} \p@renwd=\wd0 

\def\eightpoint{
  \def\rm{\fam0\eightrm}%
  \textfont0=\eightrm \scriptfont0=\sixrm \scriptscriptfont0=\fiverm%
  \textfont1=\eighti  \scriptfont1=\sixi  \scriptscriptfont1=\fivei%
  \textfont2=\eightsy \scriptfont2=\sixsy \scriptscriptfont2=\fivesy%
  \textfont\itfam=\eightit\def\it{\fam\itfam\eightit}%
  \ifprod@font
    \scriptfont\itfam=\sixit
      \scriptscriptfont\itfam=\fiveit
  \else
    \scriptfont\itfam=\eightit
      \scriptscriptfont\itfam=\eightit
  \fi
  \textfont\bffam=\eightbf%
    \scriptfont\bffam=\sixbf%
      \scriptscriptfont\bffam=\fivebf%
  \def\bf{\fam\bffam\eightbf}%
  \textfont\slfam=\eightsl\def\sl{\fam\slfam\eightsl}%
  \ifprod@font
    \scriptfont\slfam=\sixsl
      \scriptscriptfont\slfam=\fivesl
  \else
    \scriptfont\slfam=\eightsl
      \scriptscriptfont\slfam=\eightsl
  \fi
  \textfont\ttfam=\eighttt\def\tt{\fam\ttfam\eighttt}%
  \ifprod@font
    \scriptfont\ttfam=\sixtt
      \scriptscriptfont\ttfam=\fivett
  \else
    \scriptfont\ttfam=\eighttt
      \scriptscriptfont\ttfam=\eighttt
  \fi
  \textfont\scfam=\eightcsc\def\sc{\fam\scfam\eightcsc}%
  \ifprod@font
    \scriptfont\scfam=\sixcsc
      \scriptscriptfont\scfam=\fivecsc
  \else
    \scriptfont\scfam=\eightcsc
      \scriptscriptfont\scfam=\eightcsc
  \fi
  \textfont\sffam=\eightsf\def\sf{\fam\sffam\eightsf}%
  \ifprod@font
    \scriptfont\sffam=\sixsf
      \scriptscriptfont\sffam=\fivesf
  \else
    \scriptfont\sffam=\eightsf
      \scriptscriptfont\sffam=\eightsf
  \fi
  \def\oldstyle{\fam\@ne\eighti}%
  \b@ls{10pt}\rm\@viiipt%
}
\def\@viiipt{}

\def\ninepoint{
  \def\rm{\fam0\ninerm}%
  \textfont0=\ninerm \scriptfont0=\sixrm \scriptscriptfont0=\fiverm%
  \textfont1=\ninei  \scriptfont1=\sixi  \scriptscriptfont1=\fivei%
  \textfont2=\ninesy \scriptfont2=\sixsy \scriptscriptfont2=\fivesy%
  \textfont\itfam=\nineit\def\it{\fam\itfam\nineit}%
  \ifprod@font
    \scriptfont\itfam=\sixit
      \scriptscriptfont\itfam=\fiveit
  \else
    \scriptfont\itfam=\nineit
      \scriptscriptfont\itfam=\nineit
  \fi
  \textfont\bffam=\ninebf%
    \scriptfont\bffam=\sixbf%
      \scriptscriptfont\bffam=\fivebf%
  \def\bf{\fam\bffam\ninebf}%
  \textfont\slfam=\ninesl\def\sl{\fam\slfam\ninesl}%
  \ifprod@font
    \scriptfont\slfam=\sixsl
      \scriptscriptfont\slfam=\fivesl
  \else
    \scriptfont\slfam=\ninesl
      \scriptscriptfont\slfam=\ninesl
  \fi
  \textfont\ttfam=\ninett\def\tt{\fam\ttfam\ninett}%
  \ifprod@font
    \scriptfont\ttfam=\sixtt
      \scriptscriptfont\ttfam=\fivett
  \else
    \scriptfont\ttfam=\ninett
      \scriptscriptfont\ttfam=\ninett
  \fi
  \textfont\scfam=\ninecsc\def\sc{\fam\scfam\ninecsc}%
  \ifprod@font
    \scriptfont\scfam=\sixcsc
      \scriptscriptfont\scfam=\fivecsc
  \else
    \scriptfont\scfam=\ninecsc
      \scriptscriptfont\scfam=\ninecsc
  \fi
  \textfont\sffam=\ninesf\def\sf{\fam\sffam\ninesf}%
  \ifprod@font
    \scriptfont\sffam=\sixsf
      \scriptscriptfont\sffam=\fivesf
  \else
    \scriptfont\sffam=\ninesf
      \scriptscriptfont\sffam=\ninesf
  \fi
  \def\oldstyle{\fam\@ne\ninei}%
  \b@ls{\TextLeading plus \Feathering}\rm\@ixpt%
}
\def\@ixpt{}

\def\tenpoint{
  \def\rm{\fam0\tenrm}%
  \textfont0=\tenrm \scriptfont0=\sevenrm \scriptscriptfont0=\fiverm%
  \textfont1=\teni  \scriptfont1=\seveni  \scriptscriptfont1=\fivei%
  \textfont2=\tensy \scriptfont2=\sevensy \scriptscriptfont2=\fivesy%
  \textfont\itfam=\tenit\def\it{\fam\itfam\tenit}%
  \ifprod@font
    \scriptfont\itfam=\sevenit
      \scriptscriptfont\itfam=\fiveit
  \else
    \scriptfont\itfam=\tenit
      \scriptscriptfont\itfam=\tenit
  \fi
  \textfont\bffam=\tenbf%
    \scriptfont\bffam=\sevenbf%
      \scriptscriptfont\bffam=\fivebf%
  \def\bf{\fam\bffam\tenbf}%
  \textfont\slfam=\tensl\def\sl{\fam\slfam\tensl}%
  \ifprod@font
    \scriptfont\slfam=\sevensl
      \scriptscriptfont\slfam=\fivesl
  \else
    \scriptfont\slfam=\tensl
      \scriptscriptfont\slfam=\tensl
  \fi
  \textfont\ttfam=\tentt\def\tt{\fam\ttfam\tentt}%
  \ifprod@font
    \scriptfont\ttfam=\seventt
      \scriptscriptfont\ttfam=\fivett
  \else
    \scriptfont\ttfam=\tentt
      \scriptscriptfont\ttfam=\tentt
  \fi
  \textfont\scfam=\tencsc\def\sc{\fam\scfam\tencsc}%
  \ifprod@font
    \scriptfont\scfam=\sevencsc
      \scriptscriptfont\scfam=\fivecsc
  \else
    \scriptfont\scfam=\tencsc
      \scriptscriptfont\scfam=\tencsc
  \fi
  \textfont\sffam=\tensf\def\sf{\fam\sffam\tensf}%
  \ifprod@font
    \scriptfont\sffam=\sevensf
      \scriptscriptfont\sffam=\fivesf
  \else
    \scriptfont\sffam=\tensf
      \scriptscriptfont\sffam=\tensf
  \fi
  \def\oldstyle{\fam\@ne\teni}%
  \b@ls{11pt}\rm\@xpt%
}
\def\@xpt{}

\def\elevenpoint{
  \def\rm{\fam0\elevenrm}%
  \textfont0=\elevenrm \scriptfont0=\eightrm \scriptscriptfont0=\sixrm%
  \textfont1=\eleveni  \scriptfont1=\eighti  \scriptscriptfont1=\sixi%
  \textfont2=\elevensy \scriptfont2=\eightsy \scriptscriptfont2=\sixsy%
  \textfont\itfam=\elevenit\def\it{\fam\itfam\elevenit}%
  \ifprod@font
    \scriptfont\itfam=\eightit
      \scriptscriptfont\itfam=\sixit
  \else
    \scriptfont\itfam=\elevenit
      \scriptscriptfont\itfam=\elevenit
  \fi
  \textfont\bffam=\elevenbf%
    \scriptfont\bffam=\eightbf%
      \scriptscriptfont\bffam=\sixbf%
  \def\bf{\fam\bffam\elevenbf}%
  \textfont\slfam=\elevensl\def\sl{\fam\slfam\elevensl}%
  \ifprod@font
    \scriptfont\slfam=\eightsl
      \scriptscriptfont\slfam=\sixsl
  \else
    \scriptfont\slfam=\elevensl
      \scriptscriptfont\slfam=\elevensl
  \fi
  \textfont\ttfam=\eleventt\def\tt{\fam\ttfam\eleventt}%
  \ifprod@font
    \scriptfont\ttfam=\eighttt
      \scriptscriptfont\ttfam=\sixtt
  \else
    \scriptfont\ttfam=\eleventt
      \scriptscriptfont\ttfam=\eleventt
  \fi
  \textfont\scfam=\elevencsc\def\sc{\fam\scfam\elevencsc}%
  \ifprod@font
    \scriptfont\scfam=\eightcsc
      \scriptscriptfont\scfam=\sixcsc
  \else
    \scriptfont\scfam=\elevencsc
      \scriptscriptfont\scfam=\elevencsc
  \fi
  \textfont\sffam=\elevensf\def\sf{\fam\sffam\elevensf}%
  \ifprod@font
    \scriptfont\sffam=\eightsf
      \scriptscriptfont\sffam=\sixsf
  \else
    \scriptfont\sffam=\elevensf
      \scriptscriptfont\sffam=\elevensf
  \fi
  \def\oldstyle{\fam\@ne\eleveni}%
  \b@ls{13pt}\rm\@xipt%
}
\def\@xipt{}

\def\fourteenpoint{
  \def\rm{\fam0\fourteenrm}%
  \textfont0\fourteenrm  \scriptfont0\tenrm  \scriptscriptfont0\sevenrm%
  \textfont1\fourteeni   \scriptfont1\teni   \scriptscriptfont1\seveni%
  \textfont2\fourteensy  \scriptfont2\tensy  \scriptscriptfont2\sevensy%
  \textfont\itfam=\fourteenit\def\it{\fam\itfam\fourteenit}%
  \ifprod@font
    \scriptfont\itfam=\tenit
      \scriptscriptfont\itfam=\sevenit
  \else
    \scriptfont\itfam=\fourteenit
      \scriptscriptfont\itfam=\fourteenit
  \fi
  \textfont\bffam=\fourteenbf%
    \scriptfont\bffam=\tenbf%
      \scriptscriptfont\bffam=\sevenbf%
  \def\bf{\fam\bffam\fourteenbf}%
  \textfont\slfam=\fourteensl\def\sl{\fam\slfam\fourteensl}%
  \ifprod@font
    \scriptfont\slfam=\tensl
      \scriptscriptfont\slfam=\sevensl
  \else
    \scriptfont\slfam=\fourteensl
      \scriptscriptfont\slfam=\fourteensl
  \fi
  \textfont\ttfam=\fourteentt\def\tt{\fam\ttfam\fourteentt}%
  \ifprod@font
    \scriptfont\ttfam=\tentt
      \scriptscriptfont\ttfam=\seventt
  \else
    \scriptfont\ttfam=\fourteentt
      \scriptscriptfont\ttfam=\fourteentt
  \fi
  \textfont\scfam=\fourteencsc\def\sc{\fam\scfam\fourteencsc}%
  \ifprod@font
    \scriptfont\scfam=\tencsc
      \scriptscriptfont\scfam=\sevencsc
  \else
    \scriptfont\scfam=\fourteencsc
      \scriptscriptfont\scfam=\fourteencsc
  \fi
  \textfont\sffam=\fourteensf\def\sf{\fam\sffam\fourteensf}%
  \ifprod@font
    \scriptfont\sffam=\tensf
      \scriptscriptfont\sffam=\sevensf
  \else
    \scriptfont\sffam=\fourteensf
      \scriptscriptfont\sffam=\fourteensf
  \fi
  \def\oldstyle{\fam\@ne\fourteeni}%
  \b@ls{17pt}\rm\@xivpt%
}
\def\@xivpt{}

\def\seventeenpoint{
  \def\rm{\fam0\seventeenrm}%
  \textfont0\seventeenrm  \scriptfont0\twelverm  \scriptscriptfont0\tenrm%
  \textfont1\seventeeni   \scriptfont1\twelvei   \scriptscriptfont1\teni%
  \textfont2\seventeensy  \scriptfont2\twelvesy  \scriptscriptfont2\tensy%
  \textfont\itfam=\seventeenit\def\it{\fam\itfam\seventeenit}%
  \ifprod@font
    \scriptfont\itfam=\twelveit
      \scriptscriptfont\itfam=\tenit
  \else
    \scriptfont\itfam=\seventeenit
      \scriptscriptfont\itfam=\seventeenit
  \fi
  \textfont\bffam=\seventeenbf%
    \scriptfont\bffam=\twelvebf%
      \scriptscriptfont\bffam=\tenbf%
  \def\bf{\fam\bffam\seventeenbf}%
  \textfont\slfam=\seventeensl\def\sl{\fam\slfam\seventeensl}%
  \ifprod@font
    \scriptfont\slfam=\twelvesl
      \scriptscriptfont\slfam=\tensl
  \else
    \scriptfont\slfam=\seventeensl
      \scriptscriptfont\slfam=\seventeensl
  \fi
  \textfont\ttfam=\seventeentt\def\tt{\fam\ttfam\seventeentt}%
  \ifprod@font
    \scriptfont\ttfam=\twelvett
      \scriptscriptfont\ttfam=\tentt
  \else
    \scriptfont\ttfam=\seventeentt
      \scriptscriptfont\ttfam=\seventeentt
  \fi
  \textfont\scfam=\seventeencsc\def\sc{\fam\scfam\seventeencsc}%
  \ifprod@font
    \scriptfont\scfam=\twelvecsc
      \scriptscriptfont\scfam=\tencsc
  \else
    \scriptfont\scfam=\seventeencsc
      \scriptscriptfont\scfam=\seventeencsc
  \fi
  \textfont\sffam=\seventeensf\def\sf{\fam\sffam\seventeensf}%
  \ifprod@font
    \scriptfont\sffam=\twelvesf
      \scriptscriptfont\sffam=\tensf
  \else
    \scriptfont\sffam=\seventeensf
      \scriptscriptfont\sffam=\seventeensf
  \fi
  \def\oldstyle{\fam\@ne\seventeeni}%
  \b@ls{20pt}\rm\@xviipt%
}
\def\@xviipt{}

\lineskip=1pt      \normallineskip=\lineskip
\lineskiplimit=\z@ \normallineskiplimit=\lineskiplimit


\def\loadboldmathnames{%
  \def\balpha{{\bmath{\alpha}}}%
  \def\bbeta{{\bmath{\beta}}}%
  \def\bgamma{{\bmath{\gamma}}}%
  \def\bdelta{{\bmath{\delta}}}%
  \def\bepsilon{{\bmath{\epsilon}}}%
  \def\bzeta{{\bmath{\zeta}}}%
  \def\boldeta{{\bmath{\eta}}}%
  \def\btheta{{\bmath{\theta}}}%
  \def\biota{{\bmath{\iota}}}%
  \def\bkappa{{\bmath{\kappa}}}%
  \def\blambda{{\bmath{\lambda}}}%
  \def\bmu{{\bmath{\mu}}}%
  \def\bnu{{\bmath{\nu}}}%
  \def\bxi{{\bmath{\xi}}}%
  \def\bpi{{\bmath{\pi}}}%
  \def\brho{{\bmath{\rho}}}%
  \def\bsigma{{\bmath{\sigma}}}%
  \def\btau{{\bmath{\tau}}}%
  \def\bupsilon{{\bmath{\upsilon}}}%
  \def\bphi{{\bmath{\phi}}}%
  \def\bchi{{\bmath{\chi}}}%
  \def\bpsi{{\bmath{\psi}}}%
  \def\bomega{{\bmath{\omega}}}%
  \def\bvarepsilon{{\bmath{\varepsilon}}}%
  \def\bvartheta{{\bmath{\vartheta}}}%
  \def\bvarpi{{\bmath{\varpi}}}%
  \def\bvarrho{{\bmath{\varrho}}}%
  \def\bvarsigma{{\bmath{\varsigma}}}%
  \def\bvarphi{{\bmath{\varphi}}}%
  \def\baleph{{\bmath{\aleph}}}%
  \def\bimath{{\bmath{\imath}}}%
  \def\bjmath{{\bmath{\jmath}}}%
  \def\bell{{\bmath{\ell}}}%
  \def\bwp{{\bmath{\wp}}}%
  \def\bRe{{\bmath{\Re}}}%
  \def\bIm{{\bmath{\Im}}}%
  \def\bpartial{{\bmath{\partial}}}%
  \def\binfty{{\bmath{\infty}}}%
  \def\bprime{{\bmath{\prime}}}%
  \def\bemptyset{{\bmath{\emptyset}}}%
  \def\bnabla{{\bmath{\nabla}}}%
  \def\btop{{\bmath{\top}}}%
  \def\bbot{{\bmath{\bot}}}%
  \def\btriangle{{\bmath{\triangle}}}%
  \def\bforall{{\bmath{\forall}}}%
  \def\bexists{{\bmath{\exists}}}%
  \def\bneg{{\bmath{\neg}}}%
  \def\bflat{{\bmath{\flat}}}%
  \def\bnatural{{\bmath{\natural}}}%
  \def\bsharp{{\bmath{\sharp}}}%
  \def\bclubsuit{{\bmath{\clubsuit}}}%
  \def\bdiamondsuit{{\bmath{\diamondsuit}}}%
  \def\bheartsuit{{\bmath{\heartsuit}}}%
  \def\bspadesuit{{\bmath{\spadesuit}}}%
  \def\bsmallint{{\bmath{\smallint}}}%
  \def\btriangleleft{{\bmath{\triangleleft}}}%
  \def\btriangleright{{\bmath{\triangleright}}}%
  \def\bbigtriangleup{{\bmath{\bigtriangleup}}}%
  \def\bbigtriangledown{{\bmath{\bigtriangledown}}}%
  \def\bwedge{{\bmath{\wedge}}}%
  \def\bvee{{\bmath{\vee}}}%
  \def\bcap{{\bmath{\cap}}}%
  \def\bcup{{\bmath{\cup}}}%
  \def\bddagger{{\bmath{\ddagger}}}%
  \def\bdagger{{\bmath{\dagger}}}%
  \def\bsqcap{{\bmath{\sqcap}}}%
  \def\bsqcup{{\bmath{\sqcup}}}%
  \def\buplus{{\bmath{\uplus}}}%
  \def\bamalg{{\bmath{\amalg}}}%
  \def\bdiamond{{\bmath{\diamond}}}%
  \def\bbullet{{\bmath{\bullet}}}%
  \def\bwr{{\bmath{\wr}}}%
  \def\bdiv{{\bmath{\div}}}%
  \def\bodot{{\bmath{\odot}}}%
  \def\boslash{{\bmath{\oslash}}}%
  \def\botimes{{\bmath{\otimes}}}%
  \def\bominus{{\bmath{\ominus}}}%
  \def\boplus{{\bmath{\oplus}}}%
  \def\bmp{{\bmath{\mp}}}%
  \def\bpm{{\bmath{\pm}}}%
  \def\bcirc{{\bmath{\circ}}}%
  \def\bbigcirc{{\bmath{\bigcirc}}}%
  \def\bsetminus{{\bmath{\setminus}}}%
  \def\bcdot{{\bmath{\cdot}}}%
  \def\bast{{\bmath{\ast}}}%
  \def\btimes{{\bmath{\times}}}%
  \def\bstar{{\bmath{\star}}}%
  \def\bpropto{{\bmath{\propto}}}%
  \def\bsqsubseteq{{\bmath{\sqsubseteq}}}%
  \def\bsqsupseteq{{\bmath{\sqsupseteq}}}%
  \def\bparallel{{\bmath{\parallel}}}%
  \def\bmid{{\bmath{\mid}}}%
  \def\bdashv{{\bmath{\dashv}}}%
  \def\bvdash{{\bmath{\vdash}}}%
  \def\bnearrow{{\bmath{\nearrow}}}%
  \def\bsearrow{{\bmath{\searrow}}}%
  \def\bnwarrow{{\bmath{\nwarrow}}}%
  \def\bswarrow{{\bmath{\swarrow}}}%
  \def\bLeftrightarrow{{\bmath{\Leftrightarrow}}}%
  \def\bLeftarrow{{\bmath{\Leftarrow}}}%
  \def\bRightarrow{{\bmath{\Rightarrow}}}%
  \def\bleq{{\bmath{\leq}}}%
  \def\bgeq{{\bmath{\geq}}}%
  \def\bsucc{{\bmath{\succ}}}%
  \def\bprec{{\bmath{\prec}}}%
  \def\bapprox{{\bmath{\approx}}}%
  \def\bsucceq{{\bmath{\succeq}}}%
  \def\bpreceq{{\bmath{\preceq}}}%
  \def\bsupset{{\bmath{\supset}}}%
  \def\bsubset{{\bmath{\subset}}}%
  \def\bsupseteq{{\bmath{\supseteq}}}%
  \def\bsubseteq{{\bmath{\subseteq}}}%
  \def\bin{{\bmath{\in}}}%
  \def\bni{{\bmath{\ni}}}%
  \def\bgg{{\bmath{\gg}}}%
  \def\bll{{\bmath{\ll}}}%
  \def\bnot{{\bmath{\not}}}%
  \def\bleftrightarrow{{\bmath{\leftrightarrow}}}%
  \def\bleftarrow{{\bmath{\leftarrow}}}%
  \def\brightarrow{{\bmath{\rightarrow}}}%
  \def\bmapstochar{{\bmath{\mapstochar}}}%
  \def\bsim{{\bmath{\sim}}}%
  \def\bsimeq{{\bmath{\simeq}}}%
  \def\bperp{{\bmath{\perp}}}%
  \def\bequiv{{\bmath{\equiv}}}%
  \def\basymp{{\bmath{\asymp}}}%
  \def\bsmile{{\bmath{\smile}}}%
  \def\bfrown{{\bmath{\frown}}}%
  \def\bleftharpoonup{{\bmath{\leftharpoonup}}}%
  \def\bleftharpoondown{{\bmath{\leftharpoondown}}}%
  \def\brightharpoonup{{\bmath{\rightharpoonup}}}%
  \def\brightharpoondown{{\bmath{\rightharpoondown}}}%
  \def\blhook{{\bmath{\lhook}}}%
  \def\brhook{{\bmath{\rhook}}}%
  \def\bldotp{{\bmath{\ldotp}}}%
  \def\bcdotp{{\bmath{\cdotp}}}%
}

\def\,{\relax\ifmmode \mskip\thinmuskip\else \thinspace\fi}
\let\protect=\relax

\long\def\@ifundefined#1#2#3{\expandafter\ifx\csname
  #1\endcsname\relax#2\else#3\fi}




\newtoks\math@groups \math@groups={}
\def\addtom@thgroup#1#2{#1\expandafter{\the#1#2}} 



\def\addtosizeh@ok#1#2#3#4{%
  \expandafter\def\csname @#1pt\endcsname{%
    \def\s@ze{#2}\def\ss@ze{#3}\def\sss@ze{#4}\the\math@groups%
  }%
}



\let\resetsizehook=\addtosizeh@ok


\ifprod@font
  \addtosizeh@ok{viii} {8} {6}  {5}
  \addtosizeh@ok{ix}   {9} {6}  {5}
  \addtosizeh@ok{x}    {10}{7}  {5}
  \addtosizeh@ok{xi}   {11}{8}  {6}
  \addtosizeh@ok{xiv}  {14}{10} {7}
  \addtosizeh@ok{xvii} {17}{12}{10}
\else
  \addtosizeh@ok{viii} {8}     {6}     {5}
  \addtosizeh@ok{ix}   {9}     {6}     {5}
  \addtosizeh@ok{x}    {10}    {7}     {5}
  \addtosizeh@ok{xi}   {10.95} {8}     {6}
  \addtosizeh@ok{xiv}  {14.4}  {10}    {7}
  \addtosizeh@ok{xvii} {17.28} {12}    {10}
\fi

\def\get@font#1#2#3{%
  \edef\fonts@ze{\romannumeral#3}
  \edef\fontn@me{\fonts@ze#1}
  \@ifundefined{\fontn@me}%
    {
     \global\expandafter\font\csname \fontn@me\endcsname=#2 at #3pt}%
    {}%
}

\def\ass@tfont#1#2{%
  \xdef\fam@name{\csname #1\endcsname}%
  \xdef\font@name{\csname #2\endcsname}%
  \let\textfont@name\font@name
  \textfont\fam@name\textfont@name
}

\def\ass@sfont#1#2{%
  \xdef\fam@name{\csname #1\endcsname}%
  \xdef\font@name{\csname #2\endcsname}%
  \let\textfont@name\font@name
  \scriptfont\fam@name\textfont@name
}

\def\ass@ssfont#1#2{%
  \xdef\fam@name{\csname #1\endcsname}%
  \xdef\font@name{\csname #2\endcsname}%
  \let\textfont@name\font@name
  \scriptscriptfont\fam@name\textfont@name
}


\def\NewSymbolFont#1#2{%
  \expandafter\ifx\csname sym#1fam\endcsname\relax 
    \expandafter\newfam\csname sym#1fam\endcsname
    \expandafter\edef\csname sym#1fam\endcsname{\the\allocationnumber}%
    \addtom@thgroup\math@groups{%
      \get@font{#1}{#2}{\s@ze}%
      \ass@tfont{sym#1fam}{\fontn@me}%
      \get@font{#1}{#2}{\ss@ze}%
      \ass@sfont{sym#1fam}{\fontn@me}%
      \get@font{#1}{#2}{\sss@ze}%
      \ass@ssfont{sym#1fam}{\fontn@me}%
    }%
  \else
    \errmessage{Family `#1' already defined}%
  \fi
}


\def\NewMathSymbol#1#2#3#4{%
  \edef\f@mly{\expandafter\hexnumber{\csname sym#3fam\endcsname}}%
  \mathchardef#1="#2\f@mly#4\relax
}


\newif\ifd@f

\def\NewMathDelimiter#1#2#3#4#5#6{%
  \d@ftrue
  \expandafter\ifx\csname sym#3fam\endcsname\relax
    \d@ffalse \errmessage{Family `#3' is not defined}%
  \fi
  \expandafter\ifx\csname sym#5fam\endcsname\relax
    \d@ffalse \errmessage{Family `#5' is not defined}%
  \fi
  \ifd@f
    \edef\f@mly{\expandafter\hexnumber{\csname sym#3fam\endcsname}}%
    \edef\f@mlytw@{\expandafter\hexnumber{\csname sym#5fam\endcsname}}%
    \xdef#1{\delimiter"#2\f@mly #4\f@mlytw@ #6\relax}%
  \fi
}


\def\setboxz@h{\setbox\z@\hbox}
\def\wdz@{\wd\z@}
\def\boxz@{\box\z@}
\def\setbox@ne{\setbox\@ne}
\def\wd@ne{\wd\@ne}

\def\math@atom#1#2{%
   \binrel@{#1}\binrel@@{#2}}
\def\binrel@#1{\setboxz@h{\thinmuskip0mu
  \medmuskip\m@ne mu\thickmuskip\@ne mu$#1\m@th$}%
 \setbox@ne\hbox{\thinmuskip0mu\medmuskip\m@ne mu\thickmuskip
  \@ne mu${}#1{}\m@th$}%
 \setbox\tw@\hbox{\hskip\wd@ne\hskip-\wdz@}}
\def\binrel@@#1{\ifdim\wd2<\z@\mathbin{#1}\else\ifdim\wd\tw@>\z@
 \mathrel{#1}\else{#1}\fi\fi}

\def\m@thit{1}

\def\set@skchar#1{\global\expandafter\skewchar
  \csname\fontn@me\endcsname=#1\relax}

\def\NewMathAlphabet#1#2#3{%
  \def\tst{#3}%
  \ifx\tst\empty\else 
    \expandafter\gdef\csname #1@sc\endcsname{}
  \fi
  \expandafter\def\csname #1\endcsname{
    \protect\csname @#1\endcsname}%
  \expandafter\def\csname @#1\endcsname##1{
    {%
    \begingroup
      \get@font{#1}{#2}{\s@ze}%
      \@ifundefined{#1@sc}{}{\set@skchar{#3}}%
      \ass@tfont{m@thit}{\fontn@me}%
      \get@font{#1}{#2}{\ss@ze}%
      \@ifundefined{#1@sc}{}{\set@skchar{#3}}%
      \ass@sfont{m@thit}{\fontn@me}%
      \get@font{#1}{#2}{\sss@ze}%
      \@ifundefined{#1@sc}{}{\set@skchar{#3}}%
      \ass@ssfont{m@thit}{\fontn@me}%
      \math@atom{##1}{%
      \mathchoice%
        {\hbox{$\m@th\displaystyle##1$}}%
        {\hbox{$\m@th\textstyle##1$}}%
        {\hbox{$\m@th\scriptstyle##1$}}%
        {\hbox{$\m@th\scriptscriptstyle##1$}}}%
    \endgroup
    }%
  }%
}


\newif\iffirstta  \firsttatrue

\def\set@hchar#1{\global\expandafter\hyphenchar
  \csname\fontn@me\endcsname=#1\relax}

\def\NewTextAlphabet#1#2#3{%
  \iffirstta
    \global\firsttafalse
    \newfam\scratchfam
    \edef\scrt@fam{\the\allocationnumber}%
  \fi
  \def\tst{#3}%
  \ifx\tst\empty\else 
    \expandafter\gdef\csname #1@hc\endcsname{}
  \fi
  \expandafter\def\csname #1\endcsname{
    \protect\csname t@#1\endcsname}%
  \long\expandafter\def\csname t@#1\endcsname##1{
    \ifmmode
      \typeout{Warning: do not use \expandafter\string\csname #1\endcsname
        \space in math mode}\fi%
    {%
      \get@font{#1}{#2}{\s@ze}\let\t@xtfnt=\fontn@me\relax
      \@ifundefined{#1@hc}{}{\set@hchar{#3}}%
      \ass@tfont{scrt@fam}{\fontn@me}%
      \get@font{#1}{#2}{\ss@ze}%
      \@ifundefined{#1@hc}{}{\set@hchar{#3}}%
      \ass@sfont{scrt@fam}{\fontn@me}%
      \get@font{#1}{#2}{\sss@ze}%
      \@ifundefined{#1@hc}{}{\set@hchar{#3}}%
      \ass@ssfont{scrt@fam}{\fontn@me}%
      \fam\scratchfam\csname\t@xtfnt\endcsname
    ##1%
    }%
  }%
  \expandafter\def\csname #1shape
    \endcsname{\protect\csname @#1shape\endcsname}%
  \expandafter\def\csname @#1shape\endcsname{
    \ifmmode
      \typeout{Warning: do not use \expandafter\string\csname
        #1shape\endcsname \space in math mode}\fi
      \get@font{#1}{#2}{\s@ze}\let\t@xtfnt=\fontn@me\relax
      \@ifundefined{#1@hc}{}{\set@hchar{#3}}%
      \ass@tfont{scrt@fam}{\fontn@me}%
      \get@font{#1}{#2}{\ss@ze}%
      \@ifundefined{#1@hc}{}{\set@hchar{#3}}%
      \ass@sfont{scrt@fam}{\fontn@me}%
      \get@font{#1}{#2}{\sss@ze}%
      \@ifundefined{#1@hc}{}{\set@hchar{#3}}%
      \ass@ssfont{scrt@fam}{\fontn@me}%
      \fam\scratchfam\csname\t@xtfnt\endcsname
  }%
}


\ifprod@font
  \def\math@itfnt{mtmib10}
  \def\math@syfnt{mtbsy10}
\else
  \def\math@itfnt{cmmib10}
  \def\math@syfnt{cmbsy10}
\fi

\def\m@thsy{2}

\def\bmath{\protect\@bmath}
\def\@bmath#1{%
  {%
  \begingroup
    \get@font{mthit}{\math@itfnt}{\s@ze}\set@skchar{'177}%
    \ass@tfont{m@thit}{\fontn@me}%
    \get@font{mthit}{\math@itfnt}{\ss@ze}\set@skchar{'177}%
    \ass@sfont{m@thit}{\fontn@me}%
    \get@font{mthit}{\math@itfnt}{\sss@ze}\set@skchar{'177}%
    \ass@ssfont{m@thit}{\fontn@me}%
    \get@font{mthsy}{\math@syfnt}{\s@ze}\set@skchar{'60}%
    \ass@tfont{m@thsy}{\fontn@me}%
    \get@font{mthsy}{\math@syfnt}{\ss@ze}\set@skchar{'60}%
    \ass@sfont{m@thsy}{\fontn@me}%
    \get@font{mthsy}{\math@syfnt}{\sss@ze}\set@skchar{'60}%
    \ass@ssfont{m@thsy}{\fontn@me}%
    \math@atom{#1}{%
    \mathchoice%
      {\hbox{$\m@th\displaystyle#1$}}%
      {\hbox{$\m@th\textstyle#1$}}%
      {\hbox{$\m@th\scriptstyle#1$}}%
      {\hbox{$\m@th\scriptscriptstyle#1$}}}%
  \endgroup
  }%
}



\def\sun{\hbox{$\odot$}}

\def\diameter{{\ifmmode\mathchoice
{\ooalign{\hfil\hbox{$\displaystyle/$}\hfil\crcr
{\hbox{$\displaystyle\mathchar"20D$}}}}
{\ooalign{\hfil\hbox{$\textstyle/$}\hfil\crcr
{\hbox{$\textstyle\mathchar"20D$}}}}
{\ooalign{\hfil\hbox{$\scriptstyle/$}\hfil\crcr
{\hbox{$\scriptstyle\mathchar"20D$}}}}
{\ooalign{\hfil\hbox{$\scriptscriptstyle/$}\hfil\crcr
{\hbox{$\scriptscriptstyle\mathchar"20D$}}}}
\else{\ooalign{\hfil/\hfil\crcr\mathhexbox20D}}%
\fi}}

\def\sq{\ifmmode\squareforqed\else{\unskip\nobreak\hfil
\penalty50\hskip1em\null\nobreak\hfil\squareforqed
\parfillskip=0pt\finalhyphendemerits=0\endgraf}\fi}
\def\squareforqed{\hbox{\rlap{$\sqcap$}$\sqcup$}}


\def\bbbc{{\mathchoice {\setbox0=\hbox{$\displaystyle\rm C$}\hbox{\hbox
to0pt{\kern0.4\wd0\vrule height0.9\ht0\hss}\box0}}
{\setbox0=\hbox{$\textstyle\rm C$}\hbox{\hbox
to0pt{\kern0.4\wd0\vrule height0.9\ht0\hss}\box0}}
{\setbox0=\hbox{$\scriptstyle\rm C$}\hbox{\hbox
to0pt{\kern0.4\wd0\vrule height0.9\ht0\hss}\box0}}
{\setbox0=\hbox{$\scriptscriptstyle\rm C$}\hbox{\hbox
to0pt{\kern0.4\wd0\vrule height0.9\ht0\hss}\box0}}}}
\def\bbbq{{\mathchoice {\setbox0=\hbox{$\displaystyle\rm
Q$}\hbox{\raise
0.15\ht0\hbox to0pt{\kern0.4\wd0\vrule height0.8\ht0\hss}\box0}}
{\setbox0=\hbox{$\textstyle\rm Q$}\hbox{\raise
0.15\ht0\hbox to0pt{\kern0.4\wd0\vrule height0.8\ht0\hss}\box0}}
{\setbox0=\hbox{$\scriptstyle\rm Q$}\hbox{\raise
0.15\ht0\hbox to0pt{\kern0.4\wd0\vrule height0.7\ht0\hss}\box0}}
{\setbox0=\hbox{$\scriptscriptstyle\rm Q$}\hbox{\raise
0.15\ht0\hbox to0pt{\kern0.4\wd0\vrule height0.7\ht0\hss}\box0}}}}
\def\bbbt{{\mathchoice {\setbox0=\hbox{$\displaystyle\rm
T$}\hbox{\hbox to0pt{\kern0.3\wd0\vrule height0.9\ht0\hss}\box0}}
{\setbox0=\hbox{$\textstyle\rm T$}\hbox{\hbox
to0pt{\kern0.3\wd0\vrule height0.9\ht0\hss}\box0}}
{\setbox0=\hbox{$\scriptstyle\rm T$}\hbox{\hbox
to0pt{\kern0.3\wd0\vrule height0.9\ht0\hss}\box0}}
{\setbox0=\hbox{$\scriptscriptstyle\rm T$}\hbox{\hbox
to0pt{\kern0.3\wd0\vrule height0.9\ht0\hss}\box0}}}}
\def\bbbs{{\mathchoice
{\setbox0=\hbox{$\displaystyle     \rm S$}\hbox{\raise0.5\ht0\hbox
to0pt{\kern0.35\wd0\vrule height0.45\ht0\hss}\hbox
to0pt{\kern0.55\wd0\vrule height0.5\ht0\hss}\box0}}
{\setbox0=\hbox{$\textstyle        \rm S$}\hbox{\raise0.5\ht0\hbox
to0pt{\kern0.35\wd0\vrule height0.45\ht0\hss}\hbox
to0pt{\kern0.55\wd0\vrule height0.5\ht0\hss}\box0}}
{\setbox0=\hbox{$\scriptstyle      \rm S$}\hbox{\raise0.5\ht0\hbox
to0pt{\kern0.35\wd0\vrule height0.45\ht0\hss}\raise0.05\ht0\hbox
to0pt{\kern0.5\wd0\vrule height0.45\ht0\hss}\box0}}
{\setbox0=\hbox{$\scriptscriptstyle\rm S$}\hbox{\raise0.5\ht0\hbox
to0pt{\kern0.4\wd0\vrule height0.45\ht0\hss}\raise0.05\ht0\hbox
to0pt{\kern0.55\wd0\vrule height0.45\ht0\hss}\box0}}}}
\def\bbbz{{\mathchoice {\hbox{$\sf\textstyle Z\kern-0.4em Z$}}
{\hbox{$\sf\textstyle Z\kern-0.4em Z$}}
{\hbox{$\sf\scriptstyle Z\kern-0.3em Z$}}
{\hbox{$\sf\scriptscriptstyle Z\kern-0.2em Z$}}}}


\def\Nulle{0} 
\def\Afe{1}   
\def\Hae{2}   
\def\Hbe{3}   
\def\Hce{4}   
\def\Hde{5}   


\newcount\LastMac       \LastMac=\Nulle

\newskip\half      \half=5.5pt plus 1.5pt minus 2.25pt
\newskip\one       \one=11pt plus 3pt minus 5.5pt
\newskip\onehalf   \onehalf=16.5pt plus 5.5pt minus 8.25pt
\newskip\two       \two=22pt plus 5.5pt minus 11pt

\def\Half{\addvspace{\half}}
\def\One{\addvspace{\one}}
\def\OneHalf{\addvspace{\onehalf}}
\def\Two{\addvspace{\two}}

\def\Raggedright{
  \rightskip=\z@ plus \hsize\relax
}

\def\Fullout{
  \rightskip=\z@\relax
}

\def\Hang#1#2{
  \hangindent=#1%
  \hangafter=#2\relax
}


\newif\ifsp@page
\def\pagestyle#1{\csname ps@#1\endcsname}
\def\thispagestyle#1{\global\sp@pagetrue\gdef\sp@type{#1}}

\def\ps@titlepage{%
  \def\@oddhead{\eightpoint\noindent \the\CatchLine
    \ifprod@font\else\qquad Printed\ \today\qquad
      (MN plain \TeX\ macros\ v\@version)\fi \hfil}%
  \let\@evenhead=\@oddhead
  \def\@oddfoot{\eightpoint\copyright\ \@pubyear\ RAS\hfil}%
  \def\@evenfoot{\hfil\eightpoint\noindent\copyright\ \@pubyear\ RAS}%
}

\def\ps@headings{%
  \def\@oddhead{\elevenpoint\it\noindent
    \hfill\the\RightHeader\hskip1.5em\rm\folio}%
  \def\@evenhead{\elevenpoint\noindent
    \folio\hskip1.5em\it\the\LeftHeader\hfill}%
  \def\@oddfoot{\eightpoint\noindent\copyright\ \@pubyear\ RAS,
    MNRAS {\bf \@volume}, \@pagerange\hfil}%
  \def\@evenfoot{\hfil\eightpoint\copyright\ \@pubyear\ RAS,
    MNRAS {\bf \@volume}, \@pagerange}%
}

\def\ps@plate{%
  \def\@oddhead{\eightpoint\noindent\plt@cap\hfil}%
  \def\@evenhead{\eightpoint\noindent\plt@cap\hfil}%
  \def\@oddfoot{\eightpoint\noindent\copyright\ \@pubyear\ RAS,
    MNRAS {\bf \@volume}, \@pagerange\hfil}%
  \def\@evenfoot{\hfil\eightpoint\copyright\ \@pubyear\ RAS,
    MNRAS {\bf \@volume}, \@pagerange}%
}



\def\title#1{
  \bgroup
    \vbox to 8pt{\vss}%
    \seventeenpoint
    \Raggedright
    \noindent \strut{\bf #1}\par
  \egroup
}

\def\author#1{
  \bgroup
    \ifnum\LastMac=\Afe \OneHalf\else \vskip 21pt\fi
    \fourteenpoint
    \Raggedright
    \noindent \strut #1\par
    \vskip 3pt%
  \egroup
}

\def\affiliation#1{
  \bgroup
    \vskip -4pt%
    \eightpoint
    \Raggedright
    \noindent \strut {\it #1}\par
  \egroup
  \LastMac=\Afe\relax
}

\def\acceptedline#1{
  \bgroup
    \Two
    \eightpoint
    \Raggedright
    \noindent \strut #1\par
  \egroup
}

\long\def\abstract#1{%
  \bgroup
    \vskip 20pt%
    \leftskip 11pc\rightskip\z@
    \noindent{\ninebf ABSTRACT}\par
    \tenpoint
    \Fullout
    \noindent #1\par
  \egroup
}

\long\def\keywords#1{
  \bgroup
    \Half
    \leftskip 11pc\rightskip\z@
    \tenpoint
    \Fullout
    \noindent\hbox{\bf Key words:}\ #1\par
  \egroup
}


\def\maketitle{%
  \EndOpening
  \ifsinglecol \else \MakePage\fi
}


\def\pageoffset#1#2{\hoffset=#1\relax\voffset=#2\relax}


\def\@nameuse#1{\csname #1\endcsname}
\def\arabic#1{\@arabic{\@nameuse{#1}}}
\def\alph#1{\@alph{\@nameuse{#1}}}
\def\Alph#1{\@Alph{\@nameuse{#1}}}
\def\@arabic#1{\number #1}
\def\@Alph#1{\ifcase#1\or A\or B\or C\or D\else\@Ialph{#1}\fi}
\def\@Ialph#1{\ifcase#1\or \or \or \or \or E\or F\or G\or H\or I\or J\or
   K\or L\or M\or N\or O\or P\or Q\or R\or S\or T\or U\or V\or W\or X\or
   Y\or Z\else\errmessage{Counter out of range}\fi}
\def\@alph#1{\ifcase#1\or a\or b\or c\or d\else\@ialph{#1}\fi}
\def\@ialph#1{\ifcase#1\or \or \or \or \or e\or f\or g\or h\or i\or j\or
   k\or l\or m\or n\or o\or p\or q\or r\or s\or t\or u\or v\or w\or x\or y\or
   z\else\errmessage{Counter out of range}\fi}


\newcount\Eqnno
\newcount\SubEqnno

\def\theeq{\arabic{Eqnno}}
\def\thesubeq{\alph{SubEqnno}}

\def\stepeq{\relax
  \global\SubEqnno \z@
  \global\advance\Eqnno \@ne\relax
  {\rm (\theeq)}%
}

\def\startsubeq{\relax
  \global\SubEqnno \z@
  \global\advance\Eqnno \@ne\relax
  \stepsubeq
}

\def\stepsubeq{\relax
  \global\advance\SubEqnno \@ne\relax
  {\rm (\theeq\thesubeq)}%
}


\newcount\Sec        
\newcount\SecSec
\newcount\SecSecSec

\def\thesection{\arabic{Sec}}
\def\thesubsection{\thesection.\arabic{SecSec}}
\def\thesubsubsection{\thesubsection.\arabic{SecSecSec}}

\Sec=\z@

\def\:{\let\@sptoken= } \:  
\def\:{\@xifnch} \expandafter\def\: {\futurelet\@tempc\@ifnch}

\def\@ifnextchar#1#2#3{%
  \let\@tempMACe #1%
  \def\@tempMACa{#2}%
  \def\@tempMACb{#3}%
  \futurelet \@tempMACc\@ifnch%
}

\def\@ifnch{%
\ifx \@tempMACc \@sptoken%
  \let\@tempMACd\@xifnch%
\else%
  \ifx \@tempMACc \@tempMACe%
    \let\@tempMACd\@tempMACa%
  \else%
    \let\@tempMACd\@tempMACb%
  \fi%
\fi%
\@tempMACd%
}

\def\@ifstar#1#2{\@ifnextchar *{\def\@tempMACa*{#1}\@tempMACa}{#2}}

\newskip\@tempskipb

\def\addvspace#1{%
  \ifvmode\else \endgraf\fi%
  \ifdim\lastskip=\z@%
    \vskip #1\relax%
  \else%
    \@tempskipb#1\relax\@xaddvskip%
  \fi%
}

\def\@xaddvskip{%
  \ifdim\lastskip<\@tempskipb%
    \vskip-\lastskip%
    \vskip\@tempskipb\relax%
  \else%
    \ifdim\@tempskipb<\z@%
      \ifdim\lastskip<\z@ \else%
        \advance\@tempskipb\lastskip%
        \vskip-\lastskip\vskip\@tempskipb%
      \fi%
    \fi%
  \fi%
}

\newskip\@tmpSKIP

\def\addpen#1{%
  \ifvmode
    \if@nobreak
    \else
      \ifdim\lastskip=\z@
        \penalty#1\relax
      \else
        \@tmpSKIP=\lastskip
        \vskip -\lastskip
        \penalty#1\vskip\@tmpSKIP
      \fi
    \fi
  \fi
}

\newcount\@clubpen   \@clubpen=\clubpenalty
\newif\if@nobreak    \@nobreakfalse

\def\@noafterindent{%
  \global\@nobreaktrue
  \everypar{\if@nobreak
              \global\@nobreakfalse
              \clubpenalty \@M
              {\setbox\z@\lastbox}%
              \LastMac=\Nulle\relax%
            \else
              \clubpenalty \@clubpen
              \everypar{}%
            \fi}%
}

\newcount\gds@cbrk   \gds@cbrk=-300

\def\@nohdbrk{\interlinepenalty \@M\relax}

\let\@par=\par
\def\@restorepar{\def\par{\@par}}

\newif\if@endpe   \@endpefalse
 
\def\@doendpe{\@endpetrue \@nobreakfalse \LastMac=\Nulle\relax%
     \def\par{\@restorepar\everypar{}\par\@endpefalse}%
              \everypar{\setbox\z@\lastbox\everypar{}\@endpefalse}%
}

\def\section{\@ifstar{\@ssection}{\@section}}

\def\@section#1{
  \if@nobreak
    \everypar{}%
    \ifnum\LastMac=\Hae \addvspace{\half}\fi
  \else
    \addpen{\gds@cbrk}%
    \addvspace{\two}%
  \fi
  \bgroup
    \ninepoint\bf
    \Raggedright
    \global\advance\Sec \@ne
    \ifappendix
      \global\Eqnno=\z@ \global\SubEqnno=\z@\relax
      \def\ch@ck{#1}%
      \ifx\ch@ck\empty \def\c@lon{}\else\def\c@lon{:}\fi
      \noindent\@nohdbrk APPENDIX\ \thesection\c@lon\hskip 0.5em%
        \uppercase{#1}\par
    \else
      \noindent\@nohdbrk\thesection\hskip 1pc \uppercase{#1}\par
    \fi
    \global\SecSec=\z@
  \egroup
  \nobreak
  \vskip\half
  \nobreak
  \@noafterindent
  \LastMac=\Hae\relax
}

\def\@ssection#1{
  \if@nobreak
    \everypar{}%
    \ifnum\LastMac=\Hae \addvspace{\half}\fi
  \else
    \addpen{\gds@cbrk}%
    \addvspace{\two}%
  \fi
  \bgroup
    \ninepoint\bf
    \Raggedright
    \noindent\@nohdbrk\uppercase{#1}\par
  \egroup
  \nobreak
  \vskip\half
  \nobreak
  \@noafterindent
  \LastMac=\Hae\relax
}

\def\subsection{\@ifstar{\@ssubsection}{\@subsection}}

\def\@subsection#1{
  \if@nobreak
    \everypar{}%
    \ifnum\LastMac=\Hae \addvspace{1pt plus 1pt minus .5pt}\fi
  \else
    \addpen{\gds@cbrk}%
    \addvspace{\onehalf}%
  \fi
  \bgroup
    \ninepoint\bf
    \Raggedright
    \global\advance\SecSec \@ne
    \noindent\@nohdbrk\thesubsection \hskip 1pc\relax #1\par
    \global\SecSecSec=\z@
  \egroup
  \nobreak
  \vskip\half
  \nobreak
  \@noafterindent
  \LastMac=\Hbe\relax
}

\def\@ssubsection#1{
  \if@nobreak
    \everypar{}%
    \ifnum\LastMac=\Hae \addvspace{1pt plus 1pt minus .5pt}\fi
  \else
    \addpen{\gds@cbrk}%
    \addvspace{\onehalf}%
  \fi
  \bgroup
    \ninepoint\bf
    \Raggedright
    \noindent\@nohdbrk #1\par
  \egroup
  \nobreak
  \vskip\half
  \nobreak
  \@noafterindent
  \LastMac=\Hbe\relax
}

\def\subsubsection{\@ifstar{\@ssubsubsection}{\@subsubsection}}

\def\@subsubsection#1{
  \if@nobreak
    \everypar{}%
    \ifnum\LastMac=\Hbe \addvspace{1pt plus 1pt minus .5pt}\fi
  \else
    \addpen{\gds@cbrk}%
    \addvspace{\onehalf}%
  \fi
  \bgroup
    \ninepoint\it
    \Raggedright
    \global\advance\SecSecSec \@ne
    \noindent\@nohdbrk\thesubsubsection \hskip 1pc\relax #1\par
  \egroup
  \nobreak
  \vskip\half
  \nobreak
  \@noafterindent
  \LastMac=\Hce\relax
}

\def\@ssubsubsection#1{
  \if@nobreak
    \everypar{}%
    \ifnum\LastMac=\Hbe \addvspace{1pt plus 1pt minus .5pt}\fi
  \else
    \addpen{\gds@cbrk}%
    \addvspace{\onehalf}%
  \fi
  \bgroup
    \ninepoint\it
    \Raggedright
    \noindent\@nohdbrk #1\par
  \egroup
  \nobreak
  \vskip\half
  \nobreak
  \@noafterindent
  \LastMac=\Hce\relax
}

\def\paragraph#1{
  \if@nobreak
    \everypar{}%
  \else
    \addpen{\gds@cbrk}%
    \addvspace{\one}%
  \fi%
  \bgroup%
    \ninepoint\it
    \noindent #1\ \nobreak%
  \egroup
  \LastMac=\Hde\relax
  \ignorespaces
}


\newif\ifappendix

\def\appendix{%
  \global\appendixtrue
  \def\thesection{\Alph{Sec}}%
  \def\thesubsection{\thesection\arabic{SecSec}}%
  \def\theeq{\thesection\arabic{Eqnno}}%
  \Sec=\z@ \SecSec=\z@ \SecSecSec=\z@ \Eqnno=\z@ \SubEqnno=\z@\relax
}




\def\beginlist{%
  \par\if@nobreak \else\addvspace{\half}\fi%
  \bgroup%
    \ninepoint
    \let\item=\list@item%
}

\def\list@item{%
  \par\noindent\hskip 1em\relax%
  \ignorespaces%
}

\def\endlist{\par\egroup\addvspace{\half}\@doendpe}


\def\beginrefs{%
  \par
  \bgroup
    \eightpoint
    \Fullout
    \let\bibitem=\bib@item
}

\def\bib@item{%
  \par\parindent=1.5em\Hang{1.5em}{1}%
  \everypar={\Hang{1.5em}{1}\ignorespaces}%
  \noindent\ignorespaces
}

\def\endrefs{\par\egroup\@doendpe}


\newtoks\CatchLine

\def\@journal{Mon.\ Not.\ R.\ Astron.\ Soc.\ }  
\def\@pubyear{1994}        
\def\@pagerange{000--000}  
\def\@volume{000}          
\def\@microfiche{}         %

\def\pubyear#1{\gdef\@pubyear{#1}\@makecatchline}
\def\pagerange#1{\gdef\@pagerange{#1}\@makecatchline}
\def\volume#1{\gdef\@volume{#1}\@makecatchline}
\def\microfiche#1{\gdef\@microfiche{and Microfiche\ #1}\@makecatchline}

\def\@makecatchline{%
  \global\CatchLine{%
    {\rm \@journal {\bf \@volume},\ \@pagerange\ (\@pubyear)\ \@microfiche}}%
}

\@makecatchline 

\newtoks\LeftHeader
\def\shortauthor#1{
  \global\LeftHeader{#1}%
}

\newtoks\RightHeader
\def\shorttitle#1{
  \global\RightHeader{#1}%
}

\def\PageHead{
  \begingroup
    \ifsp@page
      \csname ps@\sp@type\endcsname
    \fi
    \ifodd\pageno
      \let\the@head=\@oddhead
    \else
      \let\the@head=\@evenhead
    \fi
    \vbox to \z@{\vskip-22.5\p@%
      \hbox to \PageWidth{\vbox to8.5\p@{}%
        \the@head
      }%
    \vss}%
  \endgroup
  \nointerlineskip
}

\gdef\PageFoot{%
  \nointerlineskip%
  \begingroup
  \ifsp@page
    \csname ps@\sp@type\endcsname
    \global\sp@pagefalse
  \fi
  \vbox to 22pt{\vfil%
    \hbox to \PageWidth{%
      \eightpoint\strut\noindent
      \ifodd\pageno
        \@oddfoot
      \else
        \@evenfoot
      \fi
    }%
  }%
  \endgroup
}

\def\today{%
  \number\day\space
  \ifcase\month\or January\or February\or March\or April\or May\or June\or
    July\or August\or September\or October\or November\or December\fi
  \space\number\year%
}

\def\authorcomment#1{%
  \gdef\PageFoot{%
    \nointerlineskip%
    \vbox to 20pt{\vfil%
      \hbox to \PageWidth{\elevenpoint\noindent \hfil #1 \hfil}}%
  }%
}


\newif\ifplate@page
\newbox\plt@box

\def\beginplatepage{%
  \let\plate=\plate@head
  \let\caption=\fig@caption
  \global\setbox\plt@box=\vbox\bgroup
  \TEMPDIMEN=\PageWidth 
  \hsize=\PageWidth\relax
}

\def\endplatepage{\par\egroup\global\plate@pagetrue}
\def\plate@head#1{\gdef\plt@cap{#1}}


\def\letters{%
  \gdef\folio{\ifnum\pageno<\z@ L\romannumeral-\pageno
    \else L\number\pageno \fi}%
}


\newdimen\mathindent

\global\mathindent=\z@
\global\everydisplay{\global\@dspwd=\displaywidth\displaysetup}


\def\@displaylines#1{
  {}$\displ@y\hbox{\vbox{\halign{$\@lign\hfil\displaystyle##\hfil$\crcr
  #1\crcr}}}${}%
}

\def\@eqalign#1{\null\vcenter{\openup\jot\m@th
  \ialign{\strut\hfil$\displaystyle{##}$&$\displaystyle{{}##}$\hfil
      \crcr#1\crcr}}%
}

\def\@eqalignno#1{
  \global\advance\@dspwd by -\mathindent%
  {}$\displ@y\hbox{\vbox{\halign to\@dspwd%
  {\hfil$\@lign\displaystyle{##}$\tabskip\z@skip
  &$\@lign\displaystyle{{}##}$\hfil\tabskip\centering
  &\llap{$\@lign##$}\tabskip\z@skip\crcr
  #1\crcr}}}${}%
}


\global\let\displaylines=\@displaylines
\global\let\eqalign=\@eqalign
\global\let\eqalignno=\@eqalignno
\global\let\leqalignno=\@eqalignno

\newdimen\@dspwd   \@dspwd=\z@
\newif\if@eqno
\newif\if@leqno
\newtoks\@eqn
\newtoks\@eq

\def\displaysetup#1$${\displaytest#1\eqno\eqno\displaytest}

\def\displaytest#1\eqno#2\eqno#3\displaytest{%
 \if!#3!\ldisplaytest#1\leqno\leqno\ldisplaytest
 \else\@eqnotrue\@leqnofalse\@eqn={#2}\@eq={#1}\fi
 \generaldisplay$$}

\def\ldisplaytest#1\leqno#2\leqno#3\ldisplaytest{%
\@eq={#1}%
 \if!#3!\@eqnofalse\else\@eqnotrue\@leqnotrue
  \@eqn={#2}\fi}

\def\generaldisplay{%
  \if@eqno
    \if@leqno
      \hbox to \displaywidth{\noindent
        \rlap{$\displaystyle\the\@eqn$}%
        \hskip\mathindent$\displaystyle\the\@eq$\hfil}%
    \else
      \hbox to \displaywidth{\noindent
        \hskip\mathindent
        $\displaystyle\the\@eq$\hfil$\displaystyle\the\@eqn$}%
    \fi
  \else
    \hbox to \displaywidth{\noindent
      \hskip\mathindent$\displaystyle\the\@eq$\hfil}%
  \fi
}


\def\@notice{%
  \par\Two%
  \noindent{\b@ls{11pt}\ninerm This paper has been produced using the
    Royal Astronomical Society/Blackwell Science \TeX\ macros.\par}%
}

\outer\def\bye{\@notice\par\vfill\supereject\end}


\def\start@mess{%
  Monthly notices of the RAS journal style (\@typeface)\space
    v\@version,\space \@verdate.%
}

\everyjob{\Warn{\start@mess}}



\newif\if@debug \@debugfalse  

\def\Print#1{\if@debug\immediate\write16{#1}\else \fi}
\def\Warn#1{\immediate\write16{#1}}
\def\wlog#1{}

\newcount\Iteration 

\def\Single{0} \def\Double{1}                 
\def\Figure{0} \def\Table{1}                  

\def\InStack{0}  
\def\InZoneA{1}
\def\InZoneB{2}
\def\InZoneC{3}

\newcount\TEMPCOUNT 
\newdimen\TEMPDIMEN 
\newbox\TEMPBOX     
\newbox\VOIDBOX     

\newcount\LengthOfStack 
\newcount\MaxItems      
\newcount\StackPointer
\newcount\Point         
\newcount\NextFigure    
\newcount\NextTable     
\newcount\NextItem      

\newcount\StatusStack   
\newcount\NumStack      
\newcount\TypeStack     
\newcount\SpanStack     
\newcount\BoxStack      

\newcount\ItemSTATUS    
\newcount\ItemNUMBER    
\newcount\ItemTYPE      
\newcount\ItemSPAN      
\newbox\ItemBOX         
\newdimen\ItemSIZE      

\newdimen\PageHeight    
\newdimen\TextLeading   
\newdimen\Feathering    
\newcount\LinesPerPage  
\newdimen\ColumnWidth   
\newdimen\ColumnGap     
\newdimen\PageWidth     
\newdimen\BodgeHeight   
\newcount\Leading       

\newdimen\ZoneBSize  
\newdimen\TextSize   
\newbox\ZoneABOX     
\newbox\ZoneBBOX     
\newbox\ZoneCBOX     

\newif\ifFirstSingleItem
\newif\ifFirstZoneA
\newif\ifMakePageInComplete
\newif\ifMoreFigures \MoreFiguresfalse 
\newif\ifMoreTables  \MoreTablesfalse  

\newif\ifFigInZoneB 
\newif\ifFigInZoneC 
\newif\ifTabInZoneB 
\newif\ifTabInZoneC

\newif\ifZoneAFullPage

\newbox\MidBOX    
\newbox\LeftBOX
\newbox\RightBOX
\newbox\PageBOX   

\newif\ifLeftCOL  
\LeftCOLtrue

\newdimen\ZoneBAdjust

\newcount\ItemFits
\def\Yes{1}
\def\No{2}


\MaxItems=15
\NextFigure=\z@        
\NextTable=\@ne

\BodgeHeight=6pt
\TextLeading=11pt    
\Leading=11
\Feathering=\z@      
\LinesPerPage=61     
\topskip=\TextLeading
\ColumnWidth=20pc    
\ColumnGap=2pc       

\newskip\ItemSepamount  
\ItemSepamount=\TextLeading plus \TextLeading minus 4pt

\parskip=\z@ plus .1pt
\parindent=18pt
\widowpenalty=\z@
\clubpenalty=10000
\tolerance=1500
\hbadness=1500
\abovedisplayskip=6pt plus 2pt minus 1pt
\belowdisplayskip=6pt plus 2pt minus 1pt
\abovedisplayshortskip=6pt plus 2pt minus 1pt
\belowdisplayshortskip=6pt plus 2pt minus 1pt

\frenchspacing

\ninepoint 

\PageHeight=682pt
\PageWidth=2\ColumnWidth
\advance\PageWidth by \ColumnGap

\pagestyle{headings}




\newcount\DUMMY \StatusStack=\allocationnumber
\newcount\DUMMY \newcount\DUMMY \newcount\DUMMY 
\newcount\DUMMY \newcount\DUMMY \newcount\DUMMY 
\newcount\DUMMY \newcount\DUMMY \newcount\DUMMY
\newcount\DUMMY \newcount\DUMMY \newcount\DUMMY 
\newcount\DUMMY \newcount\DUMMY \newcount\DUMMY

\newcount\DUMMY \NumStack=\allocationnumber
\newcount\DUMMY \newcount\DUMMY \newcount\DUMMY 
\newcount\DUMMY \newcount\DUMMY \newcount\DUMMY 
\newcount\DUMMY \newcount\DUMMY \newcount\DUMMY 
\newcount\DUMMY \newcount\DUMMY \newcount\DUMMY 
\newcount\DUMMY \newcount\DUMMY \newcount\DUMMY

\newcount\DUMMY \TypeStack=\allocationnumber
\newcount\DUMMY \newcount\DUMMY \newcount\DUMMY 
\newcount\DUMMY \newcount\DUMMY \newcount\DUMMY 
\newcount\DUMMY \newcount\DUMMY \newcount\DUMMY 
\newcount\DUMMY \newcount\DUMMY \newcount\DUMMY 
\newcount\DUMMY \newcount\DUMMY \newcount\DUMMY

\newcount\DUMMY \SpanStack=\allocationnumber
\newcount\DUMMY \newcount\DUMMY \newcount\DUMMY 
\newcount\DUMMY \newcount\DUMMY \newcount\DUMMY 
\newcount\DUMMY \newcount\DUMMY \newcount\DUMMY 
\newcount\DUMMY \newcount\DUMMY \newcount\DUMMY 
\newcount\DUMMY \newcount\DUMMY \newcount\DUMMY

\newbox\DUMMY   \BoxStack=\allocationnumber
\newbox\DUMMY   \newbox\DUMMY \newbox\DUMMY 
\newbox\DUMMY   \newbox\DUMMY \newbox\DUMMY 
\newbox\DUMMY   \newbox\DUMMY \newbox\DUMMY 
\newbox\DUMMY   \newbox\DUMMY \newbox\DUMMY 
\newbox\DUMMY   \newbox\DUMMY \newbox\DUMMY

\def\wlog{\immediate\write\m@ne}


\def\GetItemAll#1{%
 \GetItemSTATUS{#1}
 \GetItemNUMBER{#1}
 \GetItemTYPE{#1}
 \GetItemSPAN{#1}
 \GetItemBOX{#1}
}

\def\GetItemSTATUS#1{%
 \Point=\StatusStack
 \advance\Point by #1
 \global\ItemSTATUS=\count\Point
}

\def\GetItemNUMBER#1{%
 \Point=\NumStack
 \advance\Point by #1
 \global\ItemNUMBER=\count\Point
}

\def\GetItemTYPE#1{%
 \Point=\TypeStack
 \advance\Point by #1
 \global\ItemTYPE=\count\Point
}

\def\GetItemSPAN#1{%
 \Point\SpanStack
 \advance\Point by #1
 \global\ItemSPAN=\count\Point
}

\def\GetItemBOX#1{%
 \Point=\BoxStack
 \advance\Point by #1
 \global\setbox\ItemBOX=\vbox{\copy\Point}
 \global\ItemSIZE=\ht\ItemBOX
 \global\advance\ItemSIZE by \dp\ItemBOX
 \TEMPCOUNT=\ItemSIZE
 \divide\TEMPCOUNT by \Leading
 \divide\TEMPCOUNT by 65536
 \advance\TEMPCOUNT \@ne
 \ItemSIZE=\TEMPCOUNT pt
 \global\multiply\ItemSIZE by \Leading
}


\def\JoinStack{%
 \ifnum\LengthOfStack=\MaxItems 
  \Warn{WARNING: Stack is full...some items will be lost!}
 \else
  \Point=\StatusStack
  \advance\Point by \LengthOfStack
  \global\count\Point=\ItemSTATUS
  \Point=\NumStack
  \advance\Point by \LengthOfStack
  \global\count\Point=\ItemNUMBER
  \Point=\TypeStack
  \advance\Point by \LengthOfStack
  \global\count\Point=\ItemTYPE
  \Point\SpanStack
  \advance\Point by \LengthOfStack
  \global\count\Point=\ItemSPAN
  \Point=\BoxStack
  \advance\Point by \LengthOfStack
  \global\setbox\Point=\vbox{\copy\ItemBOX}
  \global\advance\LengthOfStack \@ne
  \ifnum\ItemTYPE=\Figure 
   \global\MoreFigurestrue
  \else
   \global\MoreTablestrue
  \fi
 \fi
}


\def\LeaveStack#1{%
 {\Iteration=#1
 \loop
 \ifnum\Iteration<\LengthOfStack
  \advance\Iteration \@ne
  \GetItemSTATUS{\Iteration}
   \advance\Point by \m@ne
   \global\count\Point=\ItemSTATUS
  \GetItemNUMBER{\Iteration}
   \advance\Point by \m@ne
   \global\count\Point=\ItemNUMBER
  \GetItemTYPE{\Iteration}
   \advance\Point by \m@ne
   \global\count\Point=\ItemTYPE
  \GetItemSPAN{\Iteration}
   \advance\Point by \m@ne
   \global\count\Point=\ItemSPAN
  \GetItemBOX{\Iteration}
   \advance\Point by \m@ne
   \global\setbox\Point=\vbox{\copy\ItemBOX}
 \repeat}
 \global\advance\LengthOfStack by \m@ne
}


\newif\ifStackNotClean

\def\CleanStack{%
 \StackNotCleantrue
 {\Iteration=\z@
  \loop
   \ifStackNotClean
    \GetItemSTATUS{\Iteration}
    \ifnum\ItemSTATUS=\InStack
     \advance\Iteration \@ne
     \else
      \LeaveStack{\Iteration}
    \fi
   \ifnum\LengthOfStack<\Iteration
    \StackNotCleanfalse
   \fi
 \repeat}
}


\def\FindItem#1#2{%
 \global\StackPointer=\m@ne 
 {\Iteration=\z@
  \loop
  \ifnum\Iteration<\LengthOfStack
   \GetItemSTATUS{\Iteration}
   \ifnum\ItemSTATUS=\InStack
    \GetItemTYPE{\Iteration}
    \ifnum\ItemTYPE=#1
     \GetItemNUMBER{\Iteration}
     \ifnum\ItemNUMBER=#2
      \global\StackPointer=\Iteration
      \Iteration=\LengthOfStack 
     \fi
    \fi
   \fi
  \advance\Iteration \@ne
 \repeat}
}


\def\FindNext{%
 \global\StackPointer=\m@ne 
 {\Iteration=\z@
  \loop
  \ifnum\Iteration<\LengthOfStack
   \GetItemSTATUS{\Iteration}
   \ifnum\ItemSTATUS=\InStack
    \GetItemTYPE{\Iteration}
   \ifnum\ItemTYPE=\Figure
    \ifMoreFigures
      \global\NextItem=\Figure
      \global\StackPointer=\Iteration
      \Iteration=\LengthOfStack 
    \fi
   \fi
   \ifnum\ItemTYPE=\Table
    \ifMoreTables
      \global\NextItem=\Table
      \global\StackPointer=\Iteration
      \Iteration=\LengthOfStack 
    \fi
   \fi
  \fi
  \advance\Iteration \@ne
 \repeat}
}


\def\ChangeStatus#1#2{%
 \Point=\StatusStack
 \advance\Point by #1
 \global\count\Point=#2
}



\def\Zone{\InZoneA}

\ZoneBAdjust=\z@

\def\MakePage{
 \global\ZoneBSize=\PageHeight
 \global\TextSize=\ZoneBSize
 \global\ZoneAFullPagefalse
 \global\topskip=\TextLeading
 \MakePageInCompletetrue
 \MoreFigurestrue
 \MoreTablestrue
 \FigInZoneBfalse
 \FigInZoneCfalse
 \TabInZoneBfalse
 \TabInZoneCfalse
 \global\FirstSingleItemtrue
 \global\FirstZoneAtrue
 \global\setbox\ZoneABOX=\box\VOIDBOX
 \global\setbox\ZoneBBOX=\box\VOIDBOX
 \global\setbox\ZoneCBOX=\box\VOIDBOX
 \loop
  \ifMakePageInComplete
 \FindNext
 \ifnum\StackPointer=\m@ne
  \NextItem=\m@ne
  \MoreFiguresfalse
  \MoreTablesfalse
 \fi
 \ifnum\NextItem=\Figure
   \FindItem{\Figure}{\NextFigure}
   \ifnum\StackPointer=\m@ne \global\MoreFiguresfalse
   \else
    \GetItemSPAN{\StackPointer}
    \ifnum\ItemSPAN=\Single \def\Zone{\InZoneB}\relax
     \ifFigInZoneC \global\MoreFiguresfalse\fi
    \else
     \def\Zone{\InZoneA}
     \ifFigInZoneB \def\Zone{\InZoneC}\fi
    \fi
   \fi
   \ifMoreFigures\Print{}\FigureItems\fi
 \fi
\ifnum\NextItem=\Table
   \FindItem{\Table}{\NextTable}
   \ifnum\StackPointer=\m@ne \global\MoreTablesfalse
   \else
    \GetItemSPAN{\StackPointer}
    \ifnum\ItemSPAN=\Single\relax
     \ifTabInZoneC \global\MoreTablesfalse\fi
    \else
     \def\Zone{\InZoneA}
     \ifTabInZoneB \def\Zone{\InZoneC}\fi
    \fi
   \fi
   \ifMoreTables\Print{}\TableItems\fi
 \fi
   \MakePageInCompletefalse 
   \ifMoreFigures\MakePageInCompletetrue\fi
   \ifMoreTables\MakePageInCompletetrue\fi
 \repeat
 \ifZoneAFullPage
  \global\TextSize=\z@
  \global\ZoneBSize=\z@
  \global\vsize=\z@\relax
  \global\topskip=\z@\relax
  \vbox to \z@{\vss}
  \eject
 \else
 \global\advance\ZoneBSize by -\ZoneBAdjust
 \global\vsize=\ZoneBSize
 \global\hsize=\ColumnWidth
 \global\ZoneBAdjust=\z@
 \ifdim\TextSize<23pt
 \Warn{}
 \Warn{* Making column fall short: TextSize=\the\TextSize *}
 \vskip-\lastskip\eject\fi
 \fi
}

\def\MakeRightCol{
 \global\TextSize=\ZoneBSize
 \MakePageInCompletetrue
 \MoreFigurestrue
 \MoreTablestrue
 \global\FirstSingleItemtrue
 \global\setbox\ZoneBBOX=\box\VOIDBOX
 \def\Zone{\InZoneB}
 \loop
  \ifMakePageInComplete
 \FindNext
 \ifnum\StackPointer=\m@ne
  \NextItem=\m@ne
  \MoreFiguresfalse
  \MoreTablesfalse
 \fi
 \ifnum\NextItem=\Figure
   \FindItem{\Figure}{\NextFigure}
   \ifnum\StackPointer=\m@ne \MoreFiguresfalse
   \else
    \GetItemSPAN{\StackPointer}
    \ifnum\ItemSPAN=\Double\relax
     \MoreFiguresfalse\fi
   \fi
   \ifMoreFigures\Print{}\FigureItems\fi
 \fi
 \ifnum\NextItem=\Table
   \FindItem{\Table}{\NextTable}
   \ifnum\StackPointer=\m@ne \MoreTablesfalse
   \else
    \GetItemSPAN{\StackPointer}
    \ifnum\ItemSPAN=\Double\relax
     \MoreTablesfalse\fi
   \fi
   \ifMoreTables\Print{}\TableItems\fi
 \fi
   \MakePageInCompletefalse 
   \ifMoreFigures\MakePageInCompletetrue\fi
   \ifMoreTables\MakePageInCompletetrue\fi
 \repeat
 \ifZoneAFullPage
  \global\TextSize=\z@
  \global\ZoneBSize=\z@
  \global\vsize=\z@\relax
  \global\topskip=\z@\relax
  \vbox to \z@{\vss}
  \eject
 \else
 \global\vsize=\ZoneBSize
 \global\hsize=\ColumnWidth
 \ifdim\TextSize<23pt
 \Warn{}
 \Warn{* Making column fall short: TextSize=\the\TextSize *}
 \vskip-\lastskip\eject\fi
\fi
}

\def\FigureItems{
 \Print{Considering...}
 \ShowItem{\StackPointer}
 \GetItemBOX{\StackPointer} 
 \GetItemSPAN{\StackPointer}
  \CheckFitInZone 
  \ifnum\ItemFits=\Yes
   \ifnum\ItemSPAN=\Single
     \ChangeStatus{\StackPointer}{\InZoneB} 
     \global\FigInZoneBtrue
     \ifFirstSingleItem
      \hbox{}\vskip-\BodgeHeight
     \global\advance\ItemSIZE by \TextLeading
     \fi
     \unvbox\ItemBOX\ItemSep
     \global\FirstSingleItemfalse
     \global\advance\TextSize by -\ItemSIZE
     \global\advance\TextSize by -\TextLeading
   \else
    \ifFirstZoneA
     \global\advance\ItemSIZE by \TextLeading
     \global\FirstZoneAfalse\fi
    \global\advance\TextSize by -\ItemSIZE
    \global\advance\TextSize by -\TextLeading
    \global\advance\ZoneBSize by -\ItemSIZE
    \global\advance\ZoneBSize by -\TextLeading
    \ifFigInZoneB\relax
     \else
     \ifdim\TextSize<3\TextLeading
     \global\ZoneAFullPagetrue
     \fi
    \fi
    \ChangeStatus{\StackPointer}{\Zone}
    \ifnum\Zone=\InZoneC \global\FigInZoneCtrue\fi
  \fi
   \Print{TextSize=\the\TextSize}
   \Print{ZoneBSize=\the\ZoneBSize}
  \global\advance\NextFigure \@ne
   \Print{This figure has been placed.}
  \else
   \Print{No space available for this figure...holding over.}
   \Print{}
   \global\MoreFiguresfalse
  \fi
}

\def\TableItems{
 \Print{Considering...}
 \ShowItem{\StackPointer}
 \GetItemBOX{\StackPointer} 
 \GetItemSPAN{\StackPointer}
  \CheckFitInZone 
  \ifnum\ItemFits=\Yes
   \ifnum\ItemSPAN=\Single
    \ChangeStatus{\StackPointer}{\InZoneB}
     \global\TabInZoneBtrue
     \ifFirstSingleItem
      \hbox{}\vskip-\BodgeHeight
     \global\advance\ItemSIZE by \TextLeading
     \fi
     \unvbox\ItemBOX\ItemSep
     \global\FirstSingleItemfalse
     \global\advance\TextSize by -\ItemSIZE
     \global\advance\TextSize by -\TextLeading
   \else
    \ifFirstZoneA
    \global\advance\ItemSIZE by \TextLeading
    \global\FirstZoneAfalse\fi
    \global\advance\TextSize by -\ItemSIZE
    \global\advance\TextSize by -\TextLeading
    \global\advance\ZoneBSize by -\ItemSIZE
    \global\advance\ZoneBSize by -\TextLeading
    \ifFigInZoneB\relax
     \else
     \ifdim\TextSize<3\TextLeading
     \global\ZoneAFullPagetrue
     \fi
    \fi
    \ChangeStatus{\StackPointer}{\Zone}
    \ifnum\Zone=\InZoneC \global\TabInZoneCtrue\fi
   \fi
  \global\advance\NextTable \@ne
   \Print{This table has been placed.}
  \else
  \Print{No space available for this table...holding over.}
   \Print{}
   \global\MoreTablesfalse
  \fi
}


\def\CheckFitInZone{%
{\advance\TextSize by -\ItemSIZE
 \advance\TextSize by -\TextLeading
 \ifFirstSingleItem
  \advance\TextSize by \TextLeading
 \fi
 \ifnum\Zone=\InZoneA\relax
  \else \advance\TextSize by -\ZoneBAdjust
 \fi
 \ifdim\TextSize<3\TextLeading \global\ItemFits=\No
 \else \global\ItemFits=\Yes\fi}
}

\def\BeginOpening{%
  \ninepoint
  \thispagestyle{titlepage}%
  \global\setbox\ItemBOX=\vbox\bgroup%
    \hsize=\PageWidth%
    \hrule height \z@
    \ifsinglecol\vskip 6pt\fi 
}

\let\begintopmatter=\BeginOpening  

\def\EndOpening{%
  \One
  \egroup
  \ifsinglecol
    \box\ItemBOX%
    \vskip\TextLeading plus 2\TextLeading
    \@noafterindent
  \else
    \ItemNUMBER=\z@%
    \ItemTYPE=\Figure
    \ItemSPAN=\Double
    \ItemSTATUS=\InStack
    \JoinStack
  \fi
}


\newif\if@here  \@herefalse

\def\no@float{\global\@heretrue}
\let\nofloat=\relax 

\def\beginfigure{%
  \@ifstar{\global\@dfloattrue \@bfigure}{\global\@dfloatfalse \@bfigure}%
}

\def\@bfigure#1{%
  \par
  \if@dfloat
    \ItemSPAN=\Double
    \TEMPDIMEN=\PageWidth
  \else
    \ItemSPAN=\Single
    \TEMPDIMEN=\ColumnWidth
  \fi
  \ifsinglecol
    \TEMPDIMEN=\PageWidth
  \else
    \ItemSTATUS=\InStack
    \ItemNUMBER=#1%
    \ItemTYPE=\Figure
  \fi
  \bgroup
    \hsize=\TEMPDIMEN
    \global\setbox\ItemBOX=\vbox\bgroup
      \eightpoint\nostb@ls{10pt}%
      \let\caption=\fig@caption
      \ifsinglecol \let\nofloat=\no@float\fi
}

\def\fig@caption#1{%
  \vskip 5.5pt plus 6pt%
  \bgroup 
    \eightpoint\nostb@ls{10pt}%
    \setbox\TEMPBOX=\hbox{#1}%
    \ifdim\wd\TEMPBOX>\TEMPDIMEN
      \noindent \unhbox\TEMPBOX\par
    \else
      \hbox to \hsize{\hfil\unhbox\TEMPBOX\hfil}%
    \fi
  \egroup
}

\def\endfigure{%
  \par\egroup 
  \egroup
  \ifsinglecol
    \if@here \midinsert\global\@herefalse\else \topinsert\fi
      \unvbox\ItemBOX
    \endinsert
  \else
    \JoinStack
    \Print{Processing source for figure \the\ItemNUMBER}%
  \fi
}


\newbox\tab@cap@box
\def\tab@caption#1{\global\setbox\tab@cap@box=\hbox{#1\par}}

\newtoks\tab@txt@toks
\long\def\tab@txt#1{\global\tab@txt@toks={#1}\global\table@txttrue}

\newif\iftable@txt  \table@txtfalse
\newif\if@dfloat    \@dfloatfalse

\def\begintable{%
  \@ifstar{\global\@dfloattrue \@btable}{\global\@dfloatfalse \@btable}%
}

\def\@btable#1{%
  \par
  \if@dfloat
    \ItemSPAN=\Double
    \TEMPDIMEN=\PageWidth
  \else
    \ItemSPAN=\Single
    \TEMPDIMEN=\ColumnWidth
  \fi
  \ifsinglecol
    \TEMPDIMEN=\PageWidth
  \else
    \ItemSTATUS=\InStack
    \ItemNUMBER=#1%
    \ItemTYPE=\Table
  \fi
  \bgroup
    \eightpoint\nostb@ls{10pt}%
    \global\setbox\ItemBOX=\vbox\bgroup
      \let\caption=\tab@caption
      \let\tabletext=\tab@txt
      \ifsinglecol \let\nofloat=\no@float\fi
}

\def\endtable{%
  \par\egroup 
  \egroup
  \setbox\TEMPBOX=\hbox to \TEMPDIMEN{%
    \eightpoint\nostb@ls{10pt}%
    \hss
    \vbox{%
      \hsize=\wd\ItemBOX
      \ifvoid\tab@cap@box
      \else
        \noindent\unhbox\tab@cap@box
        \vskip 5.5pt plus 6pt%
      \fi
      \box\ItemBOX
      \iftable@txt
        \vskip 10pt%
        \noindent\the\tab@txt@toks
        \global\table@txtfalse
      \fi
    }%
    \hss
  }%
  \ifsinglecol
    \if@here \midinsert\global\@herefalse\else \topinsert\fi
      \box\TEMPBOX
    \endinsert
  \else
    \global\setbox\ItemBOX=\box\TEMPBOX
    \JoinStack
    \Print{Processing source for table \the\ItemNUMBER}%
  \fi
}

\def\UnloadZoneA{%
\FirstZoneAtrue
 \Iteration=\z@
  \loop
   \ifnum\Iteration<\LengthOfStack
    \GetItemSTATUS{\Iteration}
    \ifnum\ItemSTATUS=\InZoneA
     \GetItemBOX{\Iteration}
     \ifFirstZoneA \vbox to \BodgeHeight{\vfil}%
     \FirstZoneAfalse\fi
     \unvbox\ItemBOX\ItemSep
     \LeaveStack{\Iteration}
     \else
     \advance\Iteration \@ne
   \fi
 \repeat
}

\def\UnloadZoneC{%
\Iteration=\z@
  \loop
   \ifnum\Iteration<\LengthOfStack
    \GetItemSTATUS{\Iteration}
    \ifnum\ItemSTATUS=\InZoneC
     \GetItemBOX{\Iteration}
     \ItemSep\unvbox\ItemBOX
     \LeaveStack{\Iteration}
     \else
     \advance\Iteration \@ne
   \fi
 \repeat
}


\def\ShowItem#1{
  {\GetItemAll{#1}
  \Print{\the#1:
  {TYPE=\ifnum\ItemTYPE=\Figure Figure\else Table\fi}
  {NUMBER=\the\ItemNUMBER}
  {SPAN=\ifnum\ItemSPAN=\Single Single\else Double\fi}
  {SIZE=\the\ItemSIZE}}}
}

\def\ShowStack{%
 \Print{}
 \Print{LengthOfStack = \the\LengthOfStack}
 \ifnum\LengthOfStack=\z@ \Print{Stack is empty}\fi
 \Iteration=\z@
 \loop
 \ifnum\Iteration<\LengthOfStack
  \ShowItem{\Iteration}
  \advance\Iteration \@ne
 \repeat
}

\def\B#1#2{%
\hbox{\vrule\kern-0.4pt\vbox to #2{%
\hrule width #1\vfill\hrule}\kern-0.4pt\vrule}
}


\newif\ifsinglecol   \singlecolfalse

\def\onecolumn{%
  \global\output={\singlecoloutput}%
  \global\hsize=\PageWidth
  \global\vsize=\PageHeight
  \global\ColumnWidth=\hsize
  \global\TextLeading=12pt
  \global\Leading=12
  \global\singlecoltrue
  \global\let\onecolumn=\relax
  \global\let\footnote=\sing@footnote
  \global\let\vfootnote=\sing@vfootnote
  \ninepoint 
  \message{(Single column)}%
}

\def\singlecoloutput{%
  \shipout\vbox{\PageHead\vbox to \PageHeight{\pagebody\vss}\PageFoot}%
  \advancepageno
  \ifplate@page
    \shipout\vbox{%
      \sp@pagetrue
      \def\sp@type{plate}%
      \global\plate@pagefalse
      \PageHead\vbox to \PageHeight{\unvbox\plt@box\vfil}\PageFoot%
    }%
    \message{[plate]}%
    \advancepageno
  \fi
  \ifnum\outputpenalty>-\@MM \else\dosupereject\fi%
}

\def\ItemSep{\vskip\ItemSepamount\relax}

\def\ItemSepbreak{\par\ifdim\lastskip<\ItemSepamount
  \removelastskip\penalty-200\ItemSep\fi%
}


\let\@@endinsert=\endinsert 

\def\endinsert{\egroup 
  \if@mid \dimen@\ht\z@ \advance\dimen@\dp\z@ \advance\dimen@12\p@
    \advance\dimen@\pagetotal \advance\dimen@-\pageshrink
    \ifdim\dimen@>\pagegoal\@midfalse\p@gefalse\fi\fi
  \if@mid \ItemSep\box\z@\ItemSepbreak
  \else\insert\topins{\penalty100 
    \splittopskip\z@skip
    \splitmaxdepth\maxdimen \floatingpenalty\z@
    \ifp@ge \dimen@\dp\z@
    \vbox to\vsize{\unvbox\z@\kern-\dimen@}
    \else \box\z@\nobreak\ItemSep\fi}\fi\endgroup%
}


\def\gobbleone#1{}
\def\gobbletwo#1#2{}
\let\footnote=\gobbletwo 
\let\vfootnote=\gobbleone

\def\sing@footnote#1{\let\@sf\empty 
  \ifhmode\edef\@sf{\spacefactor\the\spacefactor}\/\fi
  \hbox{$^{\hbox{\eightpoint #1}}$}\@sf\sing@vfootnote{#1}%
}

\def\sing@vfootnote#1{\insert\footins\bgroup\eightpoint\b@ls{9pt}%
  \interlinepenalty\interfootnotelinepenalty
  \splittopskip\ht\strutbox 
  \splitmaxdepth\dp\strutbox \floatingpenalty\@MM
  \leftskip\z@skip \rightskip\z@skip \spaceskip\z@skip \xspaceskip\z@skip
  \noindent $^{\scriptstyle\hbox{#1}}$\hskip 4pt%
    \footstrut\futurelet\next\fo@t%
}

\def\footnoterule{\kern-3\p@ \hrule height \z@ \kern 3\p@}

\skip\footins=19.5pt plus 12pt minus 1pt
\count\footins=1000
\dimen\footins=\maxdimen

\def\note#1#2{%
  \let\@sf=\empty \ifhmode\edef\@sf{\spacefactor\the\spacefactor}\/\fi
  #1\insert\footins\bgroup
    \eightpoint\b@ls{10pt}\rm
    \interlinepenalty\interfootnotelinepenalty
    \splitmaxdepth\dp\strutbox \floatingpenalty\@MM
    \leftskip\z@skip \rightskip\z@skip \spaceskip\z@skip \xspaceskip\z@skip
    \noindent\footstrut #1$\,$#2\strut\par
  \egroup
  \@sf\relax}


\def\landscape{%
  \global\TEMPDIMEN=\PageWidth
  \global\PageWidth=\PageHeight
  \global\PageHeight=\TEMPDIMEN
  \global\let\landscape=\relax
  \onecolumn
  \message{(landscape)}%
  \raggedbottom
}


\output{%
  \ifLeftCOL
    \global\setbox\LeftBOX=\vbox to \ZoneBSize{\box255\unvbox\ZoneBBOX
      \ifvoid\footins\else
        \vskip\skip\footins\unvbox\footins\fi
    }%
    \global\LeftCOLfalse
    \MakeRightCol
  \else
    \setbox\RightBOX=\vbox to \ZoneBSize{\box255\unvbox\ZoneBBOX
      \ifvoid\footins\else
        \vskip\skip\footins\unvbox\footins\fi
    }%
    \setbox\MidBOX=\hbox{\box\LeftBOX\hskip\ColumnGap\box\RightBOX}%
    \setbox\PageBOX=\vbox to \PageHeight{%
      \UnloadZoneA\box\MidBOX\UnloadZoneC}%
    \shipout\vbox{\PageHead\vbox to \PageHeight{\box\PageBOX\vss}\PageFoot}%
    \advancepageno
    \ifplate@page
      \shipout\vbox{%
        \sp@pagetrue
        \def\sp@type{plate}%
        \global\plate@pagefalse
        \PageHead\vbox to \PageHeight{\unvbox\plt@box\vfil}\PageFoot%
      }%
      \message{[plate]}%
      \advancepageno
    \fi
    \global\LeftCOLtrue
    \CleanStack
    \MakePage
  \fi
}


\Warn{\start@mess}

\newif\ifCUPmtplainloaded 
\ifprod@font
  \global\CUPmtplainloadedtrue
\fi

\def\mnmacrosloaded{} 

\catcode `\@=12 



\fi

%

\newif\ifAMStwofonts

\ifCUPmtplainloaded \else
  \NewTextAlphabet{textbfit} {cmbxti10} {}
  \NewTextAlphabet{textbfss} {cmssbx10} {}
  \NewMathAlphabet{mathbfit} {cmbxti10} {} 
  \NewMathAlphabet{mathbfss} {cmssbx10} {} 
  \ifAMStwofonts
    \NewSymbolFont{upmath} {eurm10}
    \NewSymbolFont{AMSa} {msam10}
    \NewMathSymbol{\upi}     {0}{upmath}{19}
    \NewMathSymbol{\umu}     {0}{upmath}{16}
    \NewMathSymbol{\upartial}{0}{upmath}{40}
    \NewMathSymbol{\leqslant}{3}{AMSa}{36}
    \NewMathSymbol{\geqslant}{3}{AMSa}{3E}

    \let\leq=\leqslant 
    \let\geq=\geqslant 
  \else
    \def\umu{\mu}
    \def\upi{\pi}
    \def\upartial{\partial}
  \fi
\fi


\pageoffset{-2.5pc}{0pc}

\loadboldmathnames



\pagerange{0--0}    
\pubyear{0000}
\volume{000}

\newcount\eqcount\eqcount=0
\newcount\figcount\figcount=0
\def\nextenum{\global\advance\eqcount by 1 \thenumber}
\def\nextfignum{\global\advance\figcount by 1 \thefignumber}
\def\thenumber{\the\eqcount}
\def\thefignumber{\the\figcount}
\def\nameeq#1{\xdef#1{\thenumber}\ignorespaces}
\def\namefig#1{\xdef#1{\thefignumber}\ignorespaces}
\def\n{\nextenum}
\def\nf{\nextfignum}
\def\na#1{\label{2}{\string#1}\nameeq#1}
\def\naf#1{\label{2}{\string#1}\namefig#1}

\newbox\labelbox\newdimen\prevprevdepth
\def\label#1#2{%
\setbox\labelbox%
\vbox to 0pt{\kern -#1\baselineskip%
\hbox to 0pt{\kern\hsize\kern1pt\fivei#2\hss}%
\vss}%
\ifvmode\prevprevdepth=\prevdepth
\vskip-\baselineskip\vskip-\parskip%
\vskip\prevprevdepth%
\box\labelbox%
\prevdepth=\prevprevdepth
\else\vadjust{\box\labelbox}\fi%
}

\def\offlabel{\def\label##1##2{}}

\def\Msun{{\ifmmode M_{\sun} \else $M_{\sun}$ \fi}}
\def\msun{{\ifmmode m_{\sun} \else $m_{\sun}$ \fi}}
\def\Lsun{{\ifmmode L_{\sun} \else $L_{\sun}$ \fi}}
\def\rsun{{\ifmmode r_{\sun} \else $r_{\sun}$ \fi}}
\def\Rsun{{\ifmmode R_{\sun} \else $R_{\sun}$ \fi}}

\def\PsfigVersion{1.10}
\def\setDriver{\DvipsDriver} 
\ifx\undefined\psfig\else \fi
%

\let\LaTeXAtSign=\@
\let\@=\relax
\edef\psfigRestoreAt{\catcode`\@=\number\catcode`@\relax}
\catcode`\@=11\relax
\newwrite\@unused
\def\ps@typeout#1{{\let\protect\string\immediate\write\@unused{#1}}}

\def\DvipsDriver{
	\ps@typeout{psfig/tex \PsfigVersion -dvips}
\def\PsfigSpecials{\DvipsSpecials} 	\def\ps@dir{/}
\def\ps@predir{} }
\def\OzTeXDriver{
	\ps@typeout{psfig/tex \PsfigVersion -oztex}
	\def\PsfigSpecials{\OzTeXSpecials}
	\def\ps@dir{:}
	\def\ps@predir{:}
	\catcode`\^^J=5
}


\def\figurepath{./:}

\def\DoPaths#1{\expandafter\EachPath#1\stoplist}
\def\leer{}
\def\EachPath#1:#2\stoplist{
  \ExistsFile{#1}{\SearchedFile}
  \ifx#2\leer
  \else
    \expandafter\EachPath#2\stoplist
  \fi}
%
%
\def\ps@dir{/}
\def\ExistsFile#1#2{%
   \openin1=\ps@predir#1\ps@dir#2
   \ifeof1
       \closein1
   \else
       \closein1
        \ifx\ps@founddir\leer
           \edef\ps@founddir{#1}
        \fi
   \fi}
%
%
\def\get@dir#1{%
  \def\ps@founddir{}
  \def\SearchedFile{#1}
  \DoPaths\figurepath
}

%
%
\def\@nnil{\@nil}
\def\@empty{}
\def\@psdonoop#1\@@#2#3{}
\def\@psdo#1:=#2\do#3{\edef\@psdotmp{#2}\ifx\@psdotmp\@empty \else
    \expandafter\@psdoloop#2,\@nil,\@nil\@@#1{#3}\fi}
\def\@psdoloop#1,#2,#3\@@#4#5{\def#4{#1}\ifx #4\@nnil \else
       #5\def#4{#2}\ifx #4\@nnil \else#5\@ipsdoloop #3\@@#4{#5}\fi\fi}
\def\@ipsdoloop#1,#2\@@#3#4{\def#3{#1}\ifx #3\@nnil 
       \let\@nextwhile=\@psdonoop \else
      #4\relax\let\@nextwhile=\@ipsdoloop\fi\@nextwhile#2\@@#3{#4}}
\def\@tpsdo#1:=#2\do#3{\xdef\@psdotmp{#2}\ifx\@psdotmp\@empty \else
    \@tpsdoloop#2\@nil\@nil\@@#1{#3}\fi}
\def\@tpsdoloop#1#2\@@#3#4{\def#3{#1}\ifx #3\@nnil 
       \let\@nextwhile=\@psdonoop \else
      #4\relax\let\@nextwhile=\@tpsdoloop\fi\@nextwhile#2\@@#3{#4}}
%
\ifx\undefined\fbox
\newdimen\fboxrule
\newdimen\fboxsep
\newdimen\ps@tempdima
\newbox\ps@tempboxa
\fboxsep = 3pt
\fboxrule = .4pt
\long\def\fbox#1{\leavevmode\setbox\ps@tempboxa\hbox{#1}\ps@tempdima\fboxrule
    \advance\ps@tempdima \fboxsep \advance\ps@tempdima \dp\ps@tempboxa
   \hbox{\lower \ps@tempdima\hbox
  {\vbox{\hrule height \fboxrule
          \hbox{\vrule width \fboxrule \hskip\fboxsep
          \vbox{\vskip\fboxsep \box\ps@tempboxa\vskip\fboxsep}\hskip 
                 \fboxsep\vrule width \fboxrule}
                 \hrule height \fboxrule}}}}
\fi
%
%
\newread\ps@stream
\newif\ifnot@eof       
\newif\if@noisy        
\newif\if@atend        
\newif\if@psfile       
%
%
{\catcode`\%=12\global\gdef\epsf@start{
\def\epsf@PS{PS}
\def\epsf@getbb#1{%
%
%
\openin\ps@stream=\ps@predir#1
\ifeof\ps@stream\ps@typeout{Error, File #1 not found}\else
%
%
   {\not@eoftrue \chardef\other=12
    \def\do##1{\catcode`##1=\other}\dospecials \catcode`\ =10
    \loop
       \if@psfile
	  \read\ps@stream to \epsf@fileline
       \else{
	  \obeyspaces
          \read\ps@stream to \epsf@tmp\global\let\epsf@fileline\epsf@tmp}
       \fi
       \ifeof\ps@stream\not@eoffalse\else
%
%
       \if@psfile\else
       \expandafter\epsf@test\epsf@fileline:. \\%
       \fi
%
%
          \expandafter\epsf@aux\epsf@fileline:. \\%
       \fi
   \ifnot@eof\repeat
   }\closein\ps@stream\fi}%
%
%
\long\def\epsf@test#1#2#3:#4\\{\def\epsf@testit{#1#2}
			\ifx\epsf@testit\epsf@start\else
\ps@typeout{Warning! File does not start with `\epsf@start'.  It may not be a PostScript file.}
			\fi
			\@psfiletrue} 
%
%
{\catcode`\%=12\global\let\epsf@percent=
%
%
%
\long\def\epsf@aux#1#2:#3\\{\ifx#1\epsf@percent
   \def\epsf@testit{#2}\ifx\epsf@testit\epsf@bblit
	\@atendfalse
        \epsf@atend #3 . \\%
	\if@atend	
	   \if@verbose{
		\ps@typeout{psfig: found `(atend)'; continuing search}
	   }\fi
        \else
        \epsf@grab #3 . . . \\%
        \not@eoffalse
        \global\no@bbfalse
        \fi
   \fi\fi}%
%
%
\def\epsf@grab #1 #2 #3 #4 #5\\{%
   \global\def\epsf@llx{#1}\ifx\epsf@llx\empty
      \epsf@grab #2 #3 #4 #5 .\\\else
   \global\def\epsf@lly{#2}%
   \global\def\epsf@urx{#3}\global\def\epsf@ury{#4}\fi}%
%
%
\def\epsf@atendlit{(atend)} 
\def\epsf@atend #1 #2 #3\\{%
   \def\epsf@tmp{#1}\ifx\epsf@tmp\empty
      \epsf@atend #2 #3 .\\\else
   \ifx\epsf@tmp\epsf@atendlit\@atendtrue\fi\fi}


\chardef\psletter = 11 
\chardef\other = 12

\newif \ifdebug 
\newif\ifc@mpute 
\c@mputetrue 

\let\then = \relax
\def\r@dian{pt }
\let\r@dians = \r@dian
\let\dimensionless@nit = \r@dian
\let\dimensionless@nits = \dimensionless@nit
\def\internal@nit{sp }
\let\internal@nits = \internal@nit
\newif\ifstillc@nverging
\def \Mess@ge #1{\ifdebug \then \message {#1} \fi}

{ 
	\catcode `\@ = \psletter
	\gdef \nodimen {\expandafter \n@dimen \the \dimen}
	\gdef \term #1 #2 #3%
	       {\edef \t@ {\the #1}
		\edef \t@@ {\expandafter \n@dimen \the #2\r@dian}%
		\t@rm {\t@} {\t@@} {#3}%
	       }
	\gdef \t@rm #1 #2 #3%
	       {{%
		\count 0 = 0
		\dimen 0 = 1 \dimensionless@nit
		\dimen 2 = #2\relax
		\Mess@ge {Calculating term #1 of \nodimen 2}%
		\loop
		\ifnum	\count 0 < #1
		\then	\advance \count 0 by 1
			\Mess@ge {Iteration \the \count 0 \space}%
			\Multiply \dimen 0 by {\dimen 2}%
			\Mess@ge {After multiplication, term = \nodimen 0}%
			\Divide \dimen 0 by {\count 0}%
			\Mess@ge {After division, term = \nodimen 0}%
		\repeat
		\Mess@ge {Final value for term #1 of 
				\nodimen 2 \space is \nodimen 0}%
		\xdef \Term {#3 = \nodimen 0 \r@dians}%
		\aftergroup \Term
	       }}
	\catcode `\p = \other
	\catcode `\t = \other
	\gdef \n@dimen #1pt{#1} 
}

\def \Divide #1by #2{\divide #1 by #2} 

\def \Multiply #1by #2
       {{
	\count 0 = #1\relax
	\count 2 = #2\relax
	\count 4 = 65536
	\Mess@ge {Before scaling, count 0 = \the \count 0 \space and
			count 2 = \the \count 2}%
	\ifnum	\count 0 > 32767 
	\then	\divide \count 0 by 4
		\divide \count 4 by 4
	\else	\ifnum	\count 0 < -32767
		\then	\divide \count 0 by 4
			\divide \count 4 by 4
		\else
		\fi
	\fi
	\ifnum	\count 2 > 32767 
	\then	\divide \count 2 by 4
		\divide \count 4 by 4
	\else	\ifnum	\count 2 < -32767
		\then	\divide \count 2 by 4
			\divide \count 4 by 4
		\else
		\fi
	\fi
	\multiply \count 0 by \count 2
	\divide \count 0 by \count 4
	\xdef \product {#1 = \the \count 0 \internal@nits}%
	\aftergroup \product
       }}

\def\r@duce{\ifdim\dimen0 > 90\r@dian \then   
		\multiply\dimen0 by -1
		\advance\dimen0 by 180\r@dian
		\r@duce
	    \else \ifdim\dimen0 < -90\r@dian \then  
		\advance\dimen0 by 360\r@dian
		\r@duce
		\fi
	    \fi}

\def\Sine#1%
       {{%
	\dimen 0 = #1 \r@dian
	\r@duce
	\ifdim\dimen0 = -90\r@dian \then
	   \dimen4 = -1\r@dian
	   \c@mputefalse
	\fi
	\ifdim\dimen0 = 90\r@dian \then
	   \dimen4 = 1\r@dian
	   \c@mputefalse
	\fi
	\ifdim\dimen0 = 0\r@dian \then
	   \dimen4 = 0\r@dian
	   \c@mputefalse
	\fi
	\ifc@mpute \then
		\divide\dimen0 by 180
		\dimen0=3.141592654\dimen0
		\dimen 2 = 3.1415926535897963\r@dian 
		\divide\dimen 2 by 2 
		\Mess@ge {Sin: calculating Sin of \nodimen 0}%
		\count 0 = 1 
		\dimen 2 = 1 \r@dian 
		\dimen 4 = 0 \r@dian 
		\loop
			\ifnum	\dimen 2 = 0 
			\then	\stillc@nvergingfalse 
			\else	\stillc@nvergingtrue
			\fi
			\ifstillc@nverging 
			\then	\term {\count 0} {\dimen 0} {\dimen 2}%
				\advance \count 0 by 2
				\count 2 = \count 0
				\divide \count 2 by 2
				\ifodd	\count 2 
				\then	\advance \dimen 4 by \dimen 2
				\else	\advance \dimen 4 by -\dimen 2
				\fi
		\repeat
	\fi		
			\xdef \sine {\nodimen 4}%
       }}

\def\Cosine#1{\ifx\sine\UnDefined\edef\Savesine{\relax}\else
		             \edef\Savesine{\sine}\fi
	{\dimen0=#1\r@dian\advance\dimen0 by 90\r@dian
	 \Sine{\nodimen 0}
	 \xdef\cosine{\sine}
	 \xdef\sine{\Savesine}}}	      

\def\psdraft{
	\def\@psdraft{0}
}
\def\psfull{
	\def\@psdraft{100}
}

\psfull

\newif\if@scalefirst
\def\psscalefirst{\@scalefirsttrue}
\def\psrotatefirst{\@scalefirstfalse}
\psrotatefirst

\newif\if@draftbox
\def\psnodraftbox{
	\@draftboxfalse
}
\def\psdraftbox{
	\@draftboxtrue
}
\@draftboxtrue

\newif\if@prologfile
\newif\if@postlogfile
\def\pssilent{
	\@noisyfalse
}
\def\psnoisy{
	\@noisytrue
}
\psnoisy
\newif\if@bbllx
\newif\if@bblly
\newif\if@bburx
\newif\if@bbury
\newif\if@height
\newif\if@width
\newif\if@rheight
\newif\if@rwidth
\newif\if@angle
\newif\if@clip
\newif\if@verbose
\def\@p@@sclip#1{\@cliptrue}
\newif\if@decmpr
\def\@p@@sfigure#1{\def\@p@sfile{null}\def\@p@sbbfile{null}\@decmprfalse
   \openin1=\ps@predir#1
   \ifeof1
	\closein1
	\get@dir{#1}
	\ifx\ps@founddir\leer
		\openin1=\ps@predir#1.bb
		\ifeof1
			\closein1
			\get@dir{#1.bb}
			\ifx\ps@founddir\leer
				\ps@typeout{Can't find #1 in \figurepath}
			\else
				\@decmprtrue
				\def\@p@sfile{\ps@founddir\ps@dir#1}
				\def\@p@sbbfile{\ps@founddir\ps@dir#1.bb}
			\fi
		\else
			\closein1
			\@decmprtrue
			\def\@p@sfile{#1}
			\def\@p@sbbfile{#1.bb}
		\fi
	\else
		\def\@p@sfile{\ps@founddir\ps@dir#1}
		\def\@p@sbbfile{\ps@founddir\ps@dir#1}
	\fi
   \else
	\closein1
	\def\@p@sfile{#1}
	\def\@p@sbbfile{#1}
   \fi
}
\def\@p@@sfile#1{\@p@@sfigure{#1}}
\def\@p@@sbbllx#1{
		\@bbllxtrue
		\dimen100=#1
		\edef\@p@sbbllx{\number\dimen100}
}
\def\@p@@sbblly#1{
		\@bbllytrue
		\dimen100=#1
		\edef\@p@sbblly{\number\dimen100}
}
\def\@p@@sbburx#1{
		\@bburxtrue
		\dimen100=#1
		\edef\@p@sbburx{\number\dimen100}
}
\def\@p@@sbbury#1{
		\@bburytrue
		\dimen100=#1
		\edef\@p@sbbury{\number\dimen100}
}
\def\@p@@sheight#1{
		\@heighttrue
		\dimen100=#1
   		\edef\@p@sheight{\number\dimen100}
}
\def\@p@@swidth#1{
		\@widthtrue
		\dimen100=#1
		\edef\@p@swidth{\number\dimen100}
}
\def\@p@@srheight#1{
		\@rheighttrue
		\dimen100=#1
		\edef\@p@srheight{\number\dimen100}
}
\def\@p@@srwidth#1{
		\@rwidthtrue
		\dimen100=#1
		\edef\@p@srwidth{\number\dimen100}
}
\def\@p@@sangle#1{
		\@angletrue
		\edef\@p@sangle{#1} 
}
\def\@p@@ssilent#1{ 
		\@verbosefalse
}
\def\@p@@sprolog#1{\@prologfiletrue\def\@prologfileval{#1}}
\def\@p@@spostlog#1{\@postlogfiletrue\def\@postlogfileval{#1}}
\def\@cs@name#1{\csname #1\endcsname}
\def\@setparms#1=#2,{\@cs@name{@p@@s#1}{#2}}
%
%
\def\ps@init@parms{
		\@bbllxfalse \@bbllyfalse
		\@bburxfalse \@bburyfalse
		\@heightfalse \@widthfalse
		\@rheightfalse \@rwidthfalse
		\def\@p@sbbllx{}\def\@p@sbblly{}
		\def\@p@sbburx{}\def\@p@sbbury{}
		\def\@p@sheight{}\def\@p@swidth{}
		\def\@p@srheight{}\def\@p@srwidth{}
		\def\@p@sangle{0}
		\def\@p@sfile{} \def\@p@sbbfile{}
		\def\@p@scost{10}
		\def\@sc{}
		\@prologfilefalse
		\@postlogfilefalse
		\@clipfalse
		\if@noisy
			\@verbosetrue
		\else
			\@verbosefalse
		\fi
}
%
%
\def\parse@ps@parms#1{
	 	\@psdo\@psfiga:=#1\do
		   {\expandafter\@setparms\@psfiga,}}
%
%
\newif\ifno@bb
\def\bb@missing{
	\if@verbose{
		\ps@typeout{psfig: searching \@p@sbbfile \space  for bounding box}
	}\fi
	\no@bbtrue
	\epsf@getbb{\@p@sbbfile}
        \ifno@bb \else \bb@cull\epsf@llx\epsf@lly\epsf@urx\epsf@ury\fi
}	
\def\bb@cull#1#2#3#4{
	\dimen100=#1 bp\edef\@p@sbbllx{\number\dimen100}
	\dimen100=#2 bp\edef\@p@sbblly{\number\dimen100}
	\dimen100=#3 bp\edef\@p@sbburx{\number\dimen100}
	\dimen100=#4 bp\edef\@p@sbbury{\number\dimen100}
	\no@bbfalse
}
\newdimen\p@intvaluex
\newdimen\p@intvaluey
\def\rotate@#1#2{{\dimen0=#1 sp\dimen1=#2 sp
		  \global\p@intvaluex=\cosine\dimen0
		  \dimen3=\sine\dimen1
		  \global\advance\p@intvaluex by -\dimen3
		  \global\p@intvaluey=\sine\dimen0
		  \dimen3=\cosine\dimen1
		  \global\advance\p@intvaluey by \dimen3
		  }}
\def\compute@bb{
		\no@bbfalse
		\if@bbllx \else \no@bbtrue \fi
		\if@bblly \else \no@bbtrue \fi
		\if@bburx \else \no@bbtrue \fi
		\if@bbury \else \no@bbtrue \fi
		\ifno@bb \bb@missing \fi
		\ifno@bb \ps@typeout{FATAL ERROR: no bb supplied or found}
			\no-bb-error
		\fi
		%
%
		\count203=\@p@sbburx
		\count204=\@p@sbbury
		\advance\count203 by -\@p@sbbllx
		\advance\count204 by -\@p@sbblly
		\edef\ps@bbw{\number\count203}
		\edef\ps@bbh{\number\count204}
		\if@angle 
			\Sine{\@p@sangle}\Cosine{\@p@sangle}
	        	{\dimen100=\maxdimen\xdef\r@p@sbbllx{\number\dimen100}
					    \xdef\r@p@sbblly{\number\dimen100}
			                    \xdef\r@p@sbburx{-\number\dimen100}
					    \xdef\r@p@sbbury{-\number\dimen100}}
%
                        \def\minmaxtest{
			   \ifnum\number\p@intvaluex<\r@p@sbbllx
			      \xdef\r@p@sbbllx{\number\p@intvaluex}\fi
			   \ifnum\number\p@intvaluex>\r@p@sbburx
			      \xdef\r@p@sbburx{\number\p@intvaluex}\fi
			   \ifnum\number\p@intvaluey<\r@p@sbblly
			      \xdef\r@p@sbblly{\number\p@intvaluey}\fi
			   \ifnum\number\p@intvaluey>\r@p@sbbury
			      \xdef\r@p@sbbury{\number\p@intvaluey}\fi
			   }
			\rotate@{\@p@sbbllx}{\@p@sbblly}
			\minmaxtest
			\rotate@{\@p@sbbllx}{\@p@sbbury}
			\minmaxtest
			\rotate@{\@p@sbburx}{\@p@sbblly}
			\minmaxtest
			\rotate@{\@p@sbburx}{\@p@sbbury}
			\minmaxtest
			\edef\@p@sbbllx{\r@p@sbbllx}\edef\@p@sbblly{\r@p@sbblly}
			\edef\@p@sbburx{\r@p@sbburx}\edef\@p@sbbury{\r@p@sbbury}
		\fi
		\count203=\@p@sbburx
		\count204=\@p@sbbury
		\advance\count203 by -\@p@sbbllx
		\advance\count204 by -\@p@sbblly
		\edef\@bbw{\number\count203}
		\edef\@bbh{\number\count204}
}
%
%
\def\in@hundreds#1#2#3{\count240=#2 \count241=#3
		     \count100=\count240	
		     \divide\count100 by \count241
		     \count101=\count100
		     \multiply\count101 by \count241
		     \advance\count240 by -\count101
		     \multiply\count240 by 10
		     \count101=\count240	
		     \divide\count101 by \count241
		     \count102=\count101
		     \multiply\count102 by \count241
		     \advance\count240 by -\count102
		     \multiply\count240 by 10
		     \count102=\count240	
		     \divide\count102 by \count241
		     \count200=#1\count205=0
		     \count201=\count200
			\multiply\count201 by \count100
		 	\advance\count205 by \count201
		     \count201=\count200
			\divide\count201 by 10
			\multiply\count201 by \count101
			\advance\count205 by \count201
		     \count201=\count200
			\divide\count201 by 100
			\multiply\count201 by \count102
			\advance\count205 by \count201
		     \edef\@result{\number\count205}
}
\def\compute@wfromh{
		\in@hundreds{\@p@sheight}{\@bbw}{\@bbh}
		\edef\@p@swidth{\@result}
}
\def\compute@hfromw{
	        \in@hundreds{\@p@swidth}{\@bbh}{\@bbw}
		\edef\@p@sheight{\@result}
}
\def\compute@handw{
		\if@height 
			\if@width
			\else
				\compute@wfromh
			\fi
		\else 
			\if@width
				\compute@hfromw
			\else
				\edef\@p@sheight{\@bbh}
				\edef\@p@swidth{\@bbw}
			\fi
		\fi
}
\def\compute@resv{
		\if@rheight \else \edef\@p@srheight{\@p@sheight} \fi
		\if@rwidth \else \edef\@p@srwidth{\@p@swidth} \fi
}
%
\def\compute@sizes{
	\compute@bb
	\if@scalefirst\if@angle
	\if@width
	   \in@hundreds{\@p@swidth}{\@bbw}{\ps@bbw}
	   \edef\@p@swidth{\@result}
	\fi
	\if@height
	   \in@hundreds{\@p@sheight}{\@bbh}{\ps@bbh}
	   \edef\@p@sheight{\@result}
	\fi
	\fi\fi
	\compute@handw
	\compute@resv}
\def\OzTeXSpecials{
	\special{empty.ps /@isp {true} def}
	\special{empty.ps \@p@swidth \space \@p@sheight \space
			\@p@sbbllx \space \@p@sbblly \space
			\@p@sbburx \space \@p@sbbury \space
			startTexFig \space }
	\if@clip{
		\if@verbose{
			\ps@typeout{(clip)}
		}\fi
		\special{empty.ps doclip \space }
	}\fi
	\if@angle{
		\if@verbose{
			\ps@typeout{(rotate)}
		}\fi
		\special {empty.ps \@p@sangle \space rotate \space} 
	}\fi
	\if@prologfile
	    \special{\@prologfileval \space } \fi
	\if@decmpr{
		\if@verbose{
			\ps@typeout{psfig: Compression not available
			in OzTeX version \space }
		}\fi
	}\else{
		\if@verbose{
			\ps@typeout{psfig: including \@p@sfile \space }
		}\fi
		\special{epsf=\ps@predir\@p@sfile \space }
	}\fi
	\if@postlogfile
	    \special{\@postlogfileval \space } \fi
	\special{empty.ps /@isp {false} def}
}
\def\DvipsSpecials{
	\special{ps::[begin] 	\@p@swidth \space \@p@sheight \space
			\@p@sbbllx \space \@p@sbblly \space
			\@p@sbburx \space \@p@sbbury \space
			startTexFig \space }
	\if@clip{
		\if@verbose{
			\ps@typeout{(clip)}
		}\fi
		\special{ps:: doclip \space }
	}\fi
	\if@angle
		\if@verbose{
			\ps@typeout{(clip)}
		}\fi
		\special {ps:: \@p@sangle \space rotate \space} 
	\fi
	\if@prologfile
	    \special{ps: plotfile \@prologfileval \space } \fi
	\if@decmpr{
		\if@verbose{
			\ps@typeout{psfig: including \@p@sfile.Z \space }
		}\fi
		\special{ps: plotfile "`zcat \@p@sfile.Z" \space }
	}\else{
		\if@verbose{
			\ps@typeout{psfig: including \@p@sfile \space }
		}\fi
		\special{ps: plotfile \@p@sfile \space }
	}\fi
	\if@postlogfile
	    \special{ps: plotfile \@postlogfileval \space } \fi
	\special{ps::[end] endTexFig \space }
}
%
%
\def\psfig#1{\vbox {
	%
	\ps@init@parms
	\parse@ps@parms{#1}
	\compute@sizes
	\ifnum\@p@scost<\@psdraft{
		\PsfigSpecials 
		\vbox to \@p@srheight true sp{
			\hbox to \@p@srwidth true sp{
				\hss
			}
		\vss
                \@AddLabels 
		}
	}\else{
		\if@draftbox{		
			\hbox{\fbox{\vbox to \@p@srheight sp{
			\vss
			\hbox to \@p@srwidth sp{ \hss 
			 \hss }
			\vss
			}}}
		}\else{
			\vbox to \@p@srheight sp{
			\vss
			\hbox to \@p@srwidth sp{\hss}
			\vss
			}
		}\fi

	}\fi
}}

\def\@p@@slabels#1{%
  \def\@p@slabs{#1}%
}

\newcount\@p@sh 
\newcount\@p@sw 
\newcount\@bh
\newcount\@bw
\newcount\@dy
\def\compute@labs{
  \@p@sw=\@p@swidth  \divide\@p@sw by 65536
}%

\newread\labs
\def\@AddLabels{\openin\labs=\@p@slabs\relax
\ifeof\labs\ps@typeout{(no labels)}
\else\closein\labs 
  \ps@typeout{psfig: reading labels from \@p@slabs}
  \compute@labs
  \input \@p@slabs\relax
\fi
}

\newdimen\figsizdim \def\figsiz{\dimen\figsizdim}
\newdimen\fiddle

\newbox\labelbox
\newdimen\xlab \newdimen\ylab
\newdimen\xaux \newdimen\yaux
\newdimen\yy\newdimen\xx
\newdimen\hair\hair=3pt
\def\setlabel#1#2#3#4#5{%
\setbox\labelbox\hbox{$#1$}%
\xlab.5\wd\labelbox 
\xlab#4\xlab%
\fiddle=\hair \fiddle#4\fiddle \advance \xlab by \fiddle
\ylab.5\ht\labelbox 
\ylab#5\ylab %
\fiddle=\hair \fiddle#5\fiddle \advance \ylab by \fiddle
\advance\xlab.5\wd\labelbox%
\advance\ylab.5\ht\labelbox\advance\ylab.5\dp\labelbox%
\vbox to 0pt{\yy=#3 pt 
\divide \yy by \mag%
\multiply \yy by 1010%
\multiply \yy by \the\@p@sw%
\advance\yy by 6pt 
\advance\yy by -\ylab 
\kern-\yy%
\hbox to 0pt{%
\xx=#2 pt 
\divide \xx by \mag%
\multiply \xx by 1010%
\multiply \xx by \@p@sw
\advance\xx by -\xlab
\kern\xx\box\labelbox\hss}\vss}\ifvmode\nointerlineskip\fi}

\def\mathpro{
  \special{ps: plotfile "/home/syer/text/mma.pro-2.1" \space }
}

\psfigRestoreAt
\setDriver

\let\@=\LaTeXAtSign

\newdimen\figsizdim \def\figsiz{\dimen\figsizdim}
\def\calcfigsiz#1{\figsiz=#1 \divide\figsiz by 1010 \multiply\figsiz by \mag}
\def\psfigord#1#2{\psfig{figure=#1.ps,labels=#1.lab,width=#2}}
\def\psfigcall#1#2{\centerline{\psfigord{#1}{#2}}}

\def\psfigodd#1#2#3#4{\vbox{\hbox{\hskip#3{\psfigord{#1}{#2}}}\vskip#4}}

\def\fig[#1,#2,#3,#4]{
  \calcfigsiz{#2} 
  \vbox to #3 {\vss\psfigcall{#1}{\figsiz}}
  \vskip #4
}

\long\def\oddfig[#1,#2,#3,#4,#5,#6]{
        \calcfigsiz{#2}
        \vbox to #3 {\vss\psfigodd{#1}{\figsiz}{#4}{#5}}
        \vskip #6
}

\long\def
\fourfig[#1,#2,#3,#4,#5,#6,#7,#8]{\foursepfig[#1,#2,#3,#4,#5,#6,#7,#8,1]}

\long\def\foursepfig[#1,#2,#3,#4,#5,#6,#7,#8,#9]{
        \calcfigsiz{#5} 
        \vbox to #6 {\vss\centerline
          {\vbox{\vss\psfigodd{#1}{\figsiz}{#7}{0cm}}
           \hskip #9cm
           \vbox{\vss\psfigord{#2}{\figsiz}}}
          \vskip #9cm
          \centerline
          {\vbox{\vss\psfigodd{#3}{\figsiz}{#7}{0cm}}
           \hskip #9cm
           \vbox{\vss\psfigord{#4}{\figsiz}}}
        }
        \vskip #8
}

\long\def\sidefig[#1,#2,#3,#4,#5,#6]{
        \calcfigsiz{#3}
        \vbox to #4 {\eightpoint\vss\centerline
          {\vbox{\vss\psfigodd{#1}{\figsiz}{#5}{0cm}}
           \hskip 2cm
           \vbox{\vss\psfigord{#2}{\figsiz}}}
        }
        \vskip #6
}

\def\offfig{\def\psfigord##1##2{\relax}} 

\def\refit{}\def\refbf{}
\def\ref#1,#2,#3.{{\refit #1\/}, {\refbf #2}, #3\par\noindent}
\def\aa,#1,#2.{\ref{AA},#1,#2.}
\def\acta,#1,#2.{\ref{Acta Astron.},#1,#2.}
\def\annrev,#1,#2.{\ref{ARAA},#1,#2.}
\def\aj,#1,#2.{\ref{AJ},#1,#2.}
\def\apj,#1,#2.{\ref{ApJ},#1,#2.}
\def\apjsupp,#1,#2.{\ref{ApJ\ Supp.},#1,#2.}
\def\apspsci,#1,#2.{\ref{Ap\&SS},#1,#2.}
\def\aasupp,#1,#2.{\ref{AA\ Supp.},#1,#2.}
\def\ica,#1,#2.{\ref{Icarus},#1,#2.}
\def\grg,#1,#2.{\ref{GRG},#1,#2.}
\def\jaa,#1,#2.{\ref{J.\ Astr.\ Astrophys.},#1,#2.}
\def\mnras,#1,#2.{\ref{MNRAS},#1,#2.}
\def\nat,#1,#2.{\ref{Nature},#1,#2.}
\def\pasp,#1,#2.{\ref{PASP},#1,#2.}
\def\pasj,#1,#2.{\ref{PASJ},#1,#2.}
\def\physrev,#1,#2.{\ref{Phys.\ Rev.},#1,#2.}
\def\physrevlett,#1,#2.{\ref{Phys.\ Rev.\ Lett.},#1,#2.}
\def\physrevD,#1,#2.{\ref{Phys.\ Rev.\ D},#1,#2.}
\def\procroy,#1,#2.{\ref{Proc.\ Roy.\ Soc.},#1,#2.}
\def\revmod,#1,#2.{\ref{Rev.\ Mod.\ Phys.},#1,#2.}
\def\sova,#1,#2.{\ref{SvA},#1,#2.}

\def\vec#1{{\bf
\mathchoice
{\hbox{$\displaystyle#1$}}
{\hbox{$\textstyle#1$}}
{\hbox{$\scriptstyle#1$}}
{\hbox{$\scriptscriptstyle#1$}}}}
%

\newcount\footcount \footcount=0
\def\advftncnt{\advance\footcount by1\global\footcount=\footcount}
\def\fonote#1{\relax
  \message{N.B. footnotes do not work}
  \advftncnt\footnote{$^{\the\footcount}$}{#1}
}

\def\etal{{\it et\thinspace al\/}}
\def\adhoc{{\it ad hoc\/}}
\def\ie{{i.e}}
\def\eg{{e.g}}
\def\etc{{etc}}
\def\viz{{\sl viz\/}}
\def\cf{{cf}}

\def\frac(#1/#2){\if#2
                 \else{\textstyle{#1\over#2}}
                 \fi}

\def\reff{\label{1}{ref?}} 

\edef\vbar{|}
\def\subrm#1{_{\rm #1}}
\catcode`|=\active \let|=\subrm

\message{defined subrm}

\def\mod#1{\left\vbar#1\right\vbar} 
\def\inv#1{{1\over#1}} 
\def\pd#1#2{{\upartial#1\over \upartial#2}} \def\lr#1{\left(#1\right)}
\def\td#1{{\rm d}#1}
\def\d#1#2{{\td#1\over\td#2}}
\def\const{\hbox{const}}
 \psfull

\def\halff{{\textstyle{1\over2}}}

\offlabel

\def\y{{\ifmmode {\rm y} \else y\fi}}

\begintopmatter  

\title{Dark halo mergers and the formation of a universal profile}
\author{D. Syer and Simon D.M. White}
\affiliation{ Max-Planck-Institut f\"ur Astrophysik, 85740
Garching-bei-M\"unchen, Germany.}

\shortauthor{D. Syer \& S. White}
\shorttitle{Dark halo mergers}

\acceptedline{}

\abstract{%
We argue that a universal density profile for dark matter halos arises
as a natural consequence of hierarchical structure formation. It is a
fixed point in the process of repeated mergers.  We present analytic
and numerical arguments for the emergence of a particular form of the
profile.  At small radii, the density should
vary as $r^{-\alpha}$, with $\alpha$ determined by the way the
characteristic density of halos scales with their mass. If small halos
are dense, then $\alpha$ is large. The mass-density relation can be 
related to the power spectrum of initial fluctuations, $P(k)$, through
`formation time' arguments. Early structure formation leads to steep 
cusps. For $P(k)\sim k^n$ we find $\alpha\simeq 3(3+n)/(5+n)$. The
universal profile is generated by tidal stripping of small halos as
they merge with larger objects.}

\keywords {cosmology: theory -- dark matter -- stellar dynamics --
galaxies: kinematics and dynamics}

\maketitle

\section{Introduction}

In a series of recent papers, Navarro, Frenk \& White (1995, 1996,
1997, hereafter NFW) have claimed that dark halos formed by dissipationless 
hierarchical clustering {}from gaussian initial conditions can 
be fitted by a universal density profile of the form
$$\rho(r)=\delta {R^3\over r(R+r)^2},\eqno(\n)$$
\na\nfw
where $R$ is a characteristic length scale and $\delta$ a characteristic
density.  This function fits the numerical data presented by NFW
over a radius range of about two orders of magnitude. Equally good 
fits are obtained for high mass (rich galaxy cluster) and for low mass 
(dwarf galaxy) halos, and
in cosmological models with a wide range of initial power spectra 
$P(k)$, density parameters $\Omega$, and cosmological constants
$\Lambda$. In any given cosmology the simulated halos show a 
strong correlation between $R$ and $\delta$; low mass
halos are denser than high mass halos. NFW interpret this as 
reflecting the fact that low mass halos typically form earlier.
In independent work Cole \& Lacey (1996) came to similar
conclusions based on studies of a series of cosmological simulations 
with $P\propto k^n$ and $\Omega=1$. We shall refer to the profile
of equation (\nfw) as the NFW profile.  

It will prove useful below to consider a broader family
of density profiles: we will adopt the family defined by
$$\rho(r) = \delta {R^{\alpha+\beta}\over
r^\alpha(R^{1/\gamma}+r^{1/\gamma})^{\gamma(\beta-\alpha)}}.\eqno(\n)$$
\na\rhofit
This family has been studied extensively by Zhao (1996);
subfamilies were studied by Dehnen (1995) and by Tremaine
\etal.{} (1994).  The NFW profile is of this form: an inner cusp with
logarithmic slope $\alpha=1$, an outer envelope with logarithmic slope
$\beta=3$, and a `turn over exponent' $\gamma=1$. Another
well-known example is the Hernquist (1990) profile:
$(\alpha,\beta,\gamma)=(1,4,1)$. Both NFW and Cole \& Lacey (1996) 
compared the Hernquist profile to their numerical data. Although it
fits high mass halos almost as well as the NFW profile, its more rapid
fall-off does not fit low mass halos adequately.
Detailed comparisons of both profiles with high resolution simulations of 
galaxy clusters are also described by Tormen, Bouchet \& White (1996).

In this paper we argue that a universal profile arises as a
result of repeated mergers. Violent relaxation
of a finite mass {\it isolated} system leads to a profile
with $\rho\propto r^{-4}$ as $r$ tends to infinity (Aguilar \& White 
1986, Jaffe 1987, Merritt, Tremaine \& Johnstone 1989, Barnes 
\& Hernquist 1991). Sufficiently strong violent relaxation, either
through repeated merging or through cold inhomogeneous collapse,
produces $\rho\propto r^{-3}$ over the 
regions which contain most of the mass (White 1979, Villumsen 1982,
Duncan, Farouki \& Shapiro 1983, McGlynn 1984). There are, however,
good reasons why neither of these results should apply to halo
formation through hierarchical clustering. Cosmological mergers
are not isolated, they occur continually and {}from bound orbits, 
and they usually involve objects of very different mass. These 
properties are reflected in the other simple paradigm for halo 
formation, the spherical infall model (Gunn \& Gott 1972, Fillmore 
\& Goldreich 1984, Hoffman \& Shaham 1985, White \& Zaritsky 1992).
This model predicts relatively shallow density profiles, $\rho\sim 
r^{-2}$ for the virialised regions of halos. 

The next section uses simple arguments to show why
merging in a cosmological context might give rise to a characteristic
density profile with a central cusp. The third section then carries
out some numerical experiments to test these arguments. A final
section discusses our results.

\section{Mergers and tidal disruption }
Hierarchical clustering proceeds through a continual series of
unequal mergers. Small objects typically form at earlier times
and so have higher characteristic densities. The results of Navarro,
Frenk \& White (1996) for a CDM universe can be fit by a power law 
$$\delta\propto M^{-\nu}\eqno(\n)$$
\na\gamdef
with $\nu\simeq0.33$.  The characteristic radius is given by 
$$R\propto M^{(1+\nu)/3}.\eqno(\n)$$
\na\rmeq
Simple scaling arguments using linear theory (Kaiser 1986) predict 
that if $P(k)\sim k^n$ then
$$\nu=(3+n)/2.\eqno(\n)$$
\na\scal
$\nu=0.33$ corresponds to $n=-2.3$ as expected for a CDM universe on
the relevant scales.  

Consider a large parent halo merging with a smaller satellite system.
Two important dynamical processes are responsible for the evolution 
of the system.  Dynamical friction causes the satellite's orbit to 
decay, bringing it down into the central regions. Tidal stripping
removes material {}from the satellite and adds it to the diffuse mass of
the parent.  Our claim is that the combination of these processes
leads to $\alpha\simeq 1$ in the inner cusp.
This structure is a stable fixed point in the process of repeated 
mergers, in a sense to be defined below.


\beginfigure{1}
\fig[prof,.9\hsize,.6\hsize,2\baselineskip]

\caption{{\bf Figure \nf.}
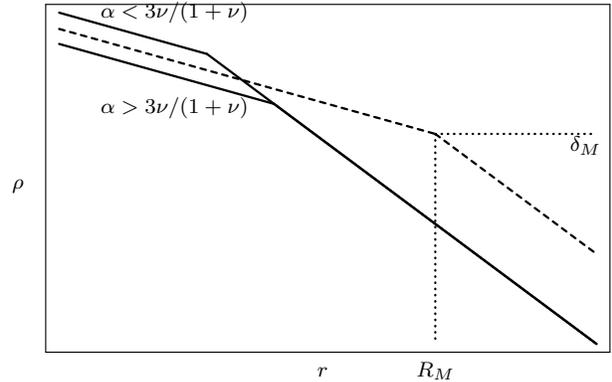
The superimposed density profiles of a parent halo of mass $M$ and
of homologous satellite halos of mass $m<M$.  The logarithmic slope of
the central cusp is $-\alpha$.  Two satellites are shown, one with
$\rho_m(R_m)<\delta_m$ (lower solid curve) and one with
$\rho_m(R_m)>\delta_m$ (upper solid curve).  The former has
$3\nu/(1+\nu)< \alpha$ and the latter $3\nu/(1+\nu) >
\alpha$.
\naf\prof }

\endfigure

A crude but useful way of understanding tidal stripping, going back at
least to the work of van H\"orner (1957), is as follows.  Consider a
satellite of mass $m$ orbiting a parent of mass $M$ at mean distance
$r$. The frequency $\Omega$ of the orbit is given roughly by
$\Omega^{2}\sim\bar\rho_M(r)$, where $\bar\rho_M(r)$ is the average
density of the parent interior to $r$.  Next consider a particle Q
orbiting within the satellite at a mean distance $\xi$ {}from its
centre. The frequency $\omega$ of this orbit is given by
$\omega^{2}\sim\bar\rho_m(\xi)$, where $\bar\rho_m(\xi)$ is the
average density of the satellite interior to $\xi$. If $\Omega=\omega$
there is a resonance between the force the satellite exerts on Q, and
the tidal force exerted by the parent. This results in a secular
transfer of energy to Q, which then escapes {}from the satellite.  Thus,
crudely speaking, the satellite is stripped of material down to radius
$\xi$ defined by $\omega(\xi)=\Omega(r)$. The newly stripped material
will orbit throughout the parent but will, on average, remain near
the orbit of the satellite at the instant that it was stripped.

Weinberg (1994abc, 1996) has shown that material can actually
be stripped off the satellite substantially within the primary
resonance $\Omega=\omega$.  The strongest effect is nevertheless
at this resonance, and since the timescale of real
cosmological mergers is not very different {}from $\Omega^{-1}$, 
there is little time for weaker effects to be felt. 
For simplicity we stick to the standard
assumption of stripping at the primary resonance, and we will use
$N$-body experiments to check our principal results.

Since clustering is hierarchical we expect the satellite to
have a higher characteristic density and a smaller characteristic
radius than the parent. As a result the density of the parent at 
the characteristic radius of the satellite $\rho_M(R_m)$ can be 
either larger or smaller than the density of the satellite 
at the same radius $\rho_m(R_m)=\delta_m$ (see
Figure~\prof{}).  (For $\alpha<3$ the average density is also
proportional to $r^{-\alpha}$.)  Thus, when dynamical friction 
has brought the satellite right to the centre of the parent, it 
will either have been entirely disrupted, or will have survived 
largely intact; the result depends on whether
$\rho_M(R_m)$ is less than or greater than
$\delta_m$.  The case of equality, where the satellite is marginally
disrupted by the parent as it reaches the centre, is given by
$$\rho_M(R_m) = \delta_M \lr{R_M\over R_m}^\alpha =
\delta_m,\eqno(\n)$$
\na\margl
which can be written in terms of the masses using equations (\rmeq)
and (\gamdef) as follows
$$1 = \lr{M\over m}^{\nu - \alpha(1+\nu)/3},\eqno(\n)$$
\na\alpre
whence
$$\alpha = {3\nu\over 1+\nu}.\eqno(\n)$$
\na\aleq
In conjunction with (\scal) this predicts 
$$\alpha=3\lr{3+n\over 5+n}.\eqno(\n)$$
\na\aln 
It is interesting that this exponent is the same as that predicted
in a completely different context, but for similar reasons, for the
small scale behaviour of mass autocorrelation function in hierarchical
clustering on the stable clustering hypothesis (Davis \& Peebles 1977).
These equations constitute our prediction for the logarithmic slope 
of the inner
cusp in cosmological halos.  For the Navarro, Frenk \& White (1996)
halos, $\nu\simeq0.33$ implying $\alpha\simeq0.75$.  This is
probably consistent with NFW's fits to these halos since their
experiments do not resolve the central cusp to $r\ll R$.
In fact $\alpha$ does not vary strongly with $n$ for cosmologically
interesting values of $-1$ to $-2.5$.
The simulations of Cole \& Lacey (1996) include the cases
$n=\{0,-1,-2\}$ corresponding to $\alpha=\{1.8,1.5,1\}$.  They do
not resolve the central cusps well, but there
is a hint in their Figure~9 that $n=0$ has steeper halos than $n=-2$.

A further argument is needed to show that halos will actually prefer
the special value $\alpha=3\nu/(1+\nu)$.  Suppose $\alpha$ in a parent
halo were smaller than this value, and it merged with a homologous
satellite.  In this case $\rho_M(R_m) < \delta_m$, so the satellite
survives intact right to the centre.  It thus boosts the density of
the parent cusp in the region $r<R_m$, and steepens the effective
logarithmic slope in $r<R_M$; {\ie.} $\alpha$ increases.  Now suppose
$\alpha$ in the parent were larger than $3\nu/(1+\nu)$.  In this case
$\rho_M(R_m) > \delta_m$, so the satellite will be disrupted at
$r>R_m$, spreading its mass out over a region which is larger than its
own characteristic radius.  It thus boosts the density of the cusp in
the region $r>R_m$, and softens the effective logarithmic slope in
$r<R_M$.  Thus hierarchical halo formation by repeated mergers should
lead to the density profile taking on a fixed point form, which
depends on the mass-density relation of merging subunits.

\beginfigure{2}
\fig[calpha,.9\hsize,.6\hsize,2\baselineskip]

\caption{{\bf Figure \nf.}
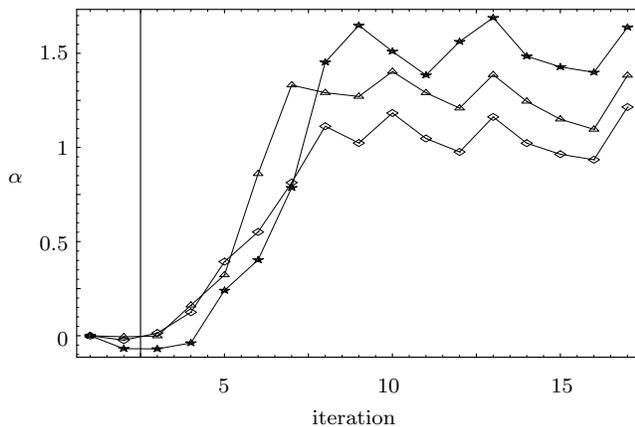
The logarithmic slope of the cusp $\alpha$ as a function of the number
of mergers in the experiments of Section~3.1. 
{}From bottom to top at the right $\nu=(0.33,0.5,0.75)$.
\naf\calpha }

\endfigure

\beginfigure{3}
\fig[ccomp,.9\hsize,.6\hsize,2\baselineskip]

\caption{{\bf Figure \nf.}
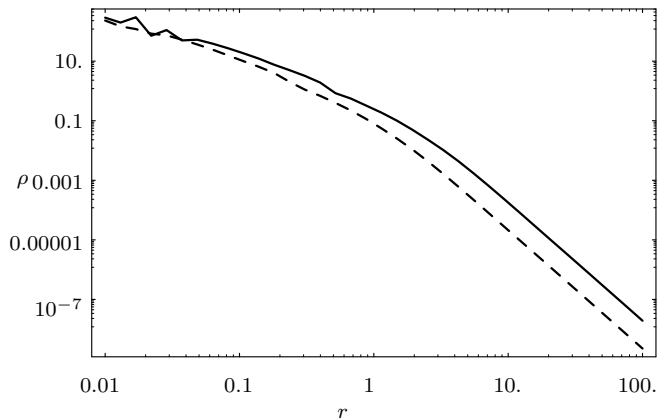
The parent halo profile {}from the last iteration in the experiment with
$(\nu,\eta,\beta)=(0.33,0,4)$ superimposed on a homologous satellite
with mass ratio $f=0.22$.  Compare Figure~\prof{}.
\naf\ccomp }

\endfigure

\begintable{1}
\caption{{\bf Table 1.} 
The results of the experiments of Section~3.1.  The input parameters are
$(\beta,\eta,\nu)$, and the output fitted profiles are specified by
$\alpha$ and $\gamma$.  The errors in the last two decimal places are
given in brackets (1$\sigma$ estimated {}from the last 10 iterations in
each case).  $\alpha|{eff}/\alpha$ is a measure of how well we resolve
the inner cusp (see text).
\null
} 
{ 
{\settabs 6 \columns

\+
\hfill $\beta$ \hfill&
\hfill $\eta$ \hfill&
\hfill $\nu$ \hfill&
\hfill $\alpha$ \hfill&
\hfill $\gamma$ \hfill&
\hfill $\alpha|{eff}/\alpha-1$ (\%) \hfill&
\cr

\+
\hfill $3$ \hfill&
\hfill $0$ \hfill&
\hfill $0.05$ \hfill&
\hfill $0.46(10)$ \hfill&
\hfill $0.57(08)$ \hfill&
\hfill $0.4$ \hfill&
\cr

\+
\hfill $3$ \hfill&
\hfill $0$ \hfill&
\hfill $0.125$ \hfill&
\hfill $0.62(08)$ \hfill&
\hfill $0.64(12)$ \hfill&
\hfill $0.7$ \hfill&
\cr

\+
\hfill $3$ \hfill&
\hfill $0$ \hfill&
\hfill $0.2$ \hfill&
\hfill $0.76(08)$ \hfill&
\hfill $0.70(13)$ \hfill&
\hfill $0.9$ \hfill&
\cr

\+
\hfill $3$ \hfill&
\hfill $0$ \hfill&
\hfill $0.33$ \hfill&
\hfill $1.00(13)$ \hfill&
\hfill $0.75(12)$ \hfill&
\hfill $0.9$ \hfill&
\cr

\+
\hfill $3$ \hfill&
\hfill $0$ \hfill&
\hfill $0.5$ \hfill&
\hfill $1.22(14)$ \hfill&
\hfill $0.80(16)$ \hfill&
\hfill $1.0$ \hfill&
\cr

\+
\hfill $3$ \hfill&
\hfill $0$ \hfill&
\hfill $0.75$ \hfill&
\hfill $1.47(14)$ \hfill&
\hfill $0.79(20)$ \hfill&
\hfill $0.9$ \hfill&
\cr

\+
\hfill $4$ \hfill&
\hfill $0$ \hfill&
\hfill $0.05$ \hfill&
\hfill $0.59(09)$ \hfill&
\hfill $0.54(03)$ \hfill&
\hfill $0.1$ \hfill&
\cr

\+
\hfill $4$ \hfill&
\hfill $0$ \hfill&
\hfill $0.125$ \hfill&
\hfill $0.81(10)$ \hfill&
\hfill $0.65(05)$ \hfill&
\hfill $0.3$ \hfill&
\cr

\+
\hfill $4$ \hfill&
\hfill $0$ \hfill&
\hfill $0.33$ \hfill&
\hfill $1.06(10)$ \hfill&
\hfill $0.83(13)$ \hfill&
\hfill $1.4$ \hfill&
\cr

\+
\hfill $4$ \hfill&
\hfill $0$ \hfill&
\hfill $0.5$ \hfill&
\hfill $1.27(10)$ \hfill&
\hfill $0.88(16)$ \hfill&
\hfill $1.5$ \hfill&
\cr

\+
\hfill $4$ \hfill&
\hfill $0$ \hfill&
\hfill $0.75$ \hfill&
\hfill $1.52(11)$ \hfill&
\hfill $0.98(16)$ \hfill&
\hfill $1.7$ \hfill&
\cr

\+
\hfill $4$ \hfill&
\hfill $-0.8$ \hfill&
\hfill $0.33$ \hfill&
\hfill $1.08(07)$ \hfill&
\hfill $0.73(07)$ \hfill&
\hfill $0.6$ \hfill&
\cr

\+
\hfill $4$ \hfill&
\hfill $-0.7$ \hfill&
\hfill $0.33$ \hfill&
\hfill $1.14(08)$ \hfill&
\hfill $0.75(05)$ \hfill&
\hfill $0.6$ \hfill&
\cr

\+
\hfill $4$ \hfill&
\hfill $-0.5$ \hfill&
\hfill $0.33$ \hfill&
\hfill $1.10(09)$ \hfill&
\hfill $0.87(06)$ \hfill&
\hfill $1.3$ \hfill&
\cr

\+
\hfill $4$ \hfill&
\hfill $0.5$ \hfill&
\hfill $0.33$ \hfill&
\hfill $1.00(09)$ \hfill&
\hfill $0.95(15)$ \hfill&
\hfill $2.9$ \hfill&
\cr

\+
\hfill $4$ \hfill&
\hfill $1.0$ \hfill&
\hfill $0.33$ \hfill&
\hfill $1.10(06)$ \hfill&
\hfill $0.76(09)$ \hfill&
\hfill $0.7$ \hfill&
\cr

\+
\hfill $4$ \hfill&
\hfill $2.0$ \hfill&
\hfill $0.33$ \hfill&
\hfill $1.12(08)$ \hfill&
\hfill $0.68(09)$ \hfill&
\hfill $0.4$ \hfill&
\cr

} 
}
\endtable

\beginfigure{4}
\fig[ga,.9\hsize,.6\hsize,2\baselineskip]

\caption{{\bf Figure \nf.}
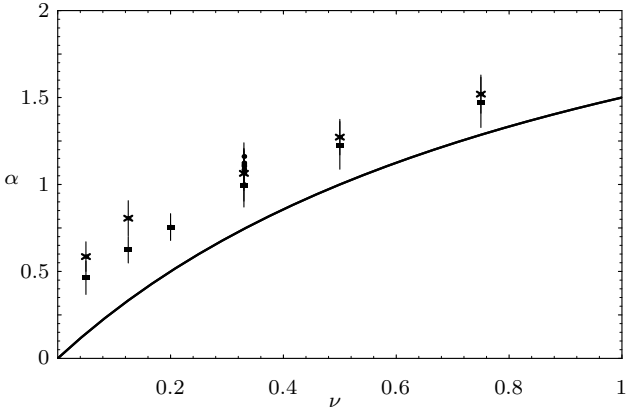
The relationship between $\nu$ and $\alpha$ as measured (with
1$\sigma$ error bars). Times signs are for $\beta=4$, squares are for
$\beta=3$ (both with $\eta=0$) and no symbol for $\beta=4$ and
different values of $\eta$ (see Table~1).  The solid line is the
prediction of equation (\aleq).
\naf\gafig }

\endfigure

\beginfigure{5}
\fig[ac,.6\hsize,.6\hsize,2\baselineskip]

\caption{{\bf Figure \nf.}
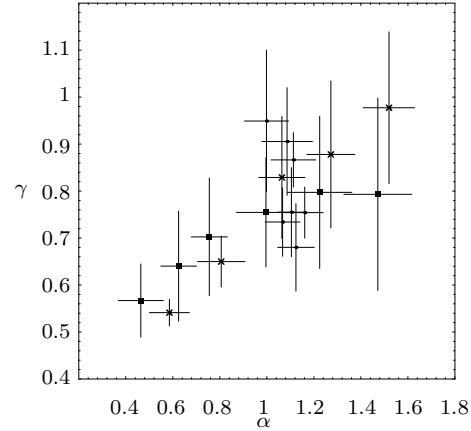
The correlation between $\gamma$ and $\alpha$ (with 1$\sigma$ error
bars).  Symbols as in Figure~\gafig{}.
\naf\acfig }

\endfigure

\section{Numerical experiments}
We now test the assertion that there is a fixed point in the process
of repeated mergers.  In outline our procedure is: to merge a halo
repeatedly with a small satellite that is created by scaling the
parent according to equation (\gamdef).  We emphasise that the
satellite is always a copy of the {\em current} halo, \ie{.} the
latest merger product.

First we take a parent
halo with mass $M$ but with an initial density profile quite different
{}from the NFW profile:
$$\rho_0(r) = \delta {R^\beta\over(R^2+r^2)^{\beta/2}},\eqno(\n)$$
\na\notnfw
with $\delta=1$, $R=1$ and $\beta$ an input parameter.  The
following steps are then carried out repeatedly:

\item{1} 
Choose a mass ratio for the next merger, $f=m/M$, by taking a random
value {}from a distribution between $f|{min}=10^{-3/(1+\nu)}$ and
$f|{max}=0.4$ in which the number of daughters with mass ratios in
$(f,f+\td{f})$ is specified by $$\td{N}=f^\eta
\td{f}.\eqno(\n)$$\na\etaeq Thus $\eta=0$ gives a uniform distribution of
daughter masses, and $\eta<0$ produces a larger number of low mass
daughters.

\item{2} Create a satellite of mass $f M$ by cloning the parent
homologously: $\rho\to \rho f^{-\nu}$, $r\to r f^{(1+\nu)/3}$.

\item{3} Merge the satellite with the parent.  Section~3.1 describes a
semi-analytic model of the merging process based on the arguments in
Section~2.  Section~3.2 describes an $N$-body implementation of the
merging process.

\subsection{Semi-analytic models}
First we implement our simplified picture of tidal stripping as
described in Section~2.  On a
logarithmic grid in $r$ between $r=0.01$ and $r=100$, a merger is
simulated as follows. Work inwards {}from the outer edge of the parent,
removing all the mass {}from the satellite where $\rho_m(\xi)=\rho_M(r)$
and placing it in the radial bin at $r$.  If the profiles cross
($\rho_M(R_m) < \delta_m$) add the satellite bin by bin to the parent,
as if it survived intact to the centre.

\noindent Figure~\calpha{} shows the
evolution of the profile as the iterations proceed for runs with
$(\eta,\beta)=(0,4)$ and $\nu=(0.33,0.5,0.75)$. Clearly the
homogeneous core is steepened and settles down into a cusp, as it was
argued it should in the previous section.  A steady state is reached,
which, as predicted, automatically flattens itself if it gets too
steep.  The final value of $\alpha$,
averaged over 10 iterations, is summarised in Table~1 and in
Figure~\gafig{}.  As shown by Figure~\gafig{}, the value of $\alpha$
is systematically larger than predicted by equation
(\aleq).  Figure~\gafig{} shows the results of experiments with
$\beta=3$ and $\beta=4$.  The lower value of $\beta$ tends to produce
slightly lower values of $\alpha$, but still consistent with the
$\beta=4$ results at the 1$\sigma$ level.

Figure~\ccomp{} shows the profile of a satellite superimposed on its
parent, so it is the equivalent of Figure~\prof{}. Since $\alpha$ is
slightly greater than predicted by equation (\aleq), the satellite is
less dense than its parent at small radii.  Note that the inner cusp
is very close to a power law.  However, Figure~\acfig{} shows that the
slope $\alpha$ is correlated with the turn over exponent $\gamma$.
This could mean that we are not resolving the power law in the cusp.
To test this we compare the measured value of $\alpha$ with the {\em
actual} logarithmic slope of the profile at the inner grid point:
$$\alpha|{eff}=
-\d{\ln\rho}{\ln r} = {\alpha+\beta r^{1/\gamma}\over1+r^{1/\gamma}}.
\eqno(\n)$$
\na\aeff
The last column of Table~1 shows that the difference between $\alpha$
and $\alpha|{eff}$ is small, thus we have adequately resolved the
central power law.

Calculations were performed with $(\beta,\nu)=(4,0.33)$ and $f$ drawn
{}from different distributions.  We adjust the relative numbers of large
and small mergers by changing the value of $\eta$ in equation
(\etaeq).  The results are summarised in Table~1, and also appear in
Figures~\acfig{} and \gafig{}.  Generally the resulting values of
$\alpha$ are similar, show no trend with $\eta$, and are statistically
consistent with each other.  There is nothing in the argument leading
to equation (\aleq) which depends on the mass ratio, so this is
expected.  The measured $\gamma$ also shows no distinct trend with
$\eta$, although, as can be seen {}from Figure~\acfig{}, varying $\eta$
can lead to large scatter in the value of $\gamma$.  All these values
are, however, consistent with each other at the 1$\sigma$ level.

\beginfigure{6}
\fig[rquants,.9\hsize,.6\hsize,2\baselineskip]

\caption{{\bf Figure \nf.}
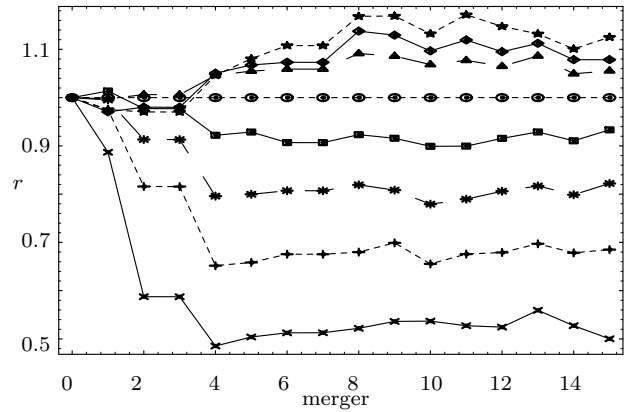
{}From bottom to top the ratio of the $10-50$ and $80,70,60$\%-\penalty
10mass radii to the half-mass radius of the halo in simulation A as a
function of the number of mergers.  Each quantity is normalised to its
initial value.
\naf\rquantsfig }

\endfigure

\beginfigure{7}
\fig[nfit,.9\hsize,.6\hsize,2\baselineskip]

\caption{{\bf Figure \nf.}
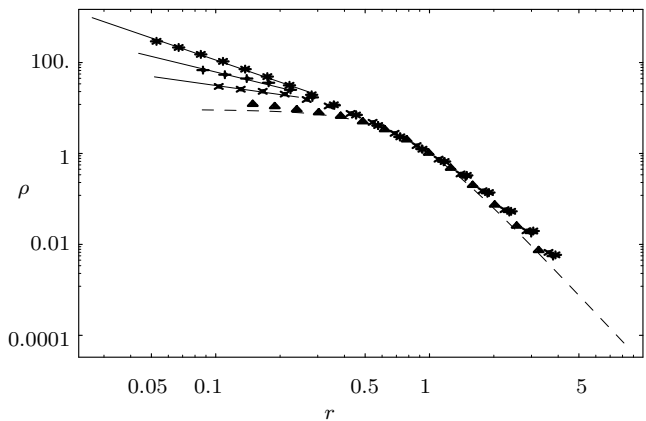
The average profile of the last 7 merger products of simulation A
(crosses), B (plus signs) and C1 (stars) with a fit to the inner parts
overlaid.  The dashed line is a fit to the original profile measured
in the same way (the fit parameters match a Plummer model well).  The
triangles close to the dotted line are a sanity check on the $N$-body
code (see text).
\naf\nfitfig }

\endfigure

\beginfigure{8}
\fig[rhs,.9\hsize,.9\hsize,2\baselineskip]

\caption{{\bf Figure \nf.}
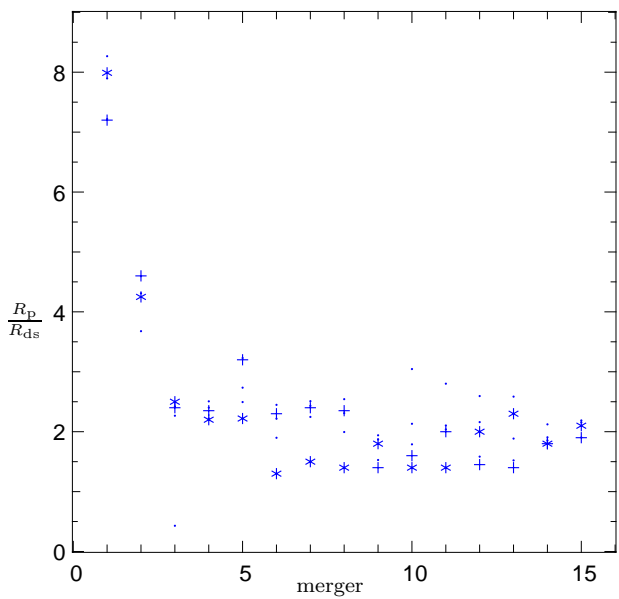
The ratio of $r_{1/2}$ in the parent halo to that of the disrupted
satellite particles {\em after} the merger, as a function of the
number of mergers, in simulations A (pluses), B (stars) and C[123]
(dots).
\naf\rhsfig }

\endfigure

\subsection{$N$-body experiments}
The code we used was the publicly available adaptive P$^3$M code of
Couchman, Pearce and Thomas (1996) (configured to simulate isolated,
collisionless systems).  Compared with these $N$-body
experiments, the simulations of Section~3.1 have better resolution and
they are much cheaper to perform.

An initial model of a Plummer sphere ($\beta=5$ in equation \notnfw)
is set up with $8000$ particles.  The satellite is placed at the
$90\%$ mass radius of the parent.  Its orbit is such that it is at
apocentre, and it has an angular momentum $\kappa$ times that of the
circular orbit with the same energy.  For $\kappa=0.3$, the apocentre
(pericentre) of the satellite orbit is thus approximately $4R$
$(0.1R)$, where $R$ is the scale radius of the parent Plummer sphere.
The satellite mass is drawn randomly according to equation (\etaeq)
with $(f|{min},f|{max},\eta)=(.05,.2,0)$.  The different values of
$\nu$ and $\kappa$ used are summarised in Table~2.  The satellite
is produced by randomly eliminating a fraction $(1-f)$ of the
particles of the parent and then scaling the positions and velocities
of the remaining particles. The smoothing
length used in the $N$-body code is a fixed fraction of the simulation
box, which is not a fixed fraction of the halo scale length.  The
largest value it took in any of the simulations was one third of the
radius of the innermost point at which the density was measured.  In
simulation A it was at most an eighth of this radius.  Simulation A
was also re-run with the smoothing length 10 times larger---the
result was not significantly different.

\begintable{2}
\caption{{\bf Table 2.} 
The parameters of the $N$-body simulations in Section~3.2.  The input
parameters are $(\nu,\kappa)$, and the output fitted profiles are
specified by $\alpha$.  The last column is the value of $\alpha$
predicted by equation (\aleq).
\null
} 
{ 
{\settabs 5 \columns

\+
\hfill Model \hfill&
\hfill $\nu$ \hfill&
\hfill $\kappa$ \hfill&
\hfill $\alpha$ \hfill&
\hfill $3\nu/(1+\nu)$&
\cr

\+
\hfill A \hfill&
\hfill $0.33$ \hfill&
\hfill $0.3$ \hfill&
\hfill $0.66$ \hfill&
\hfill $0.75$ \hfill&
\cr

\+
\hfill B \hfill&
\hfill $0.75$ \hfill&
\hfill $0.3$ \hfill&
\hfill $1.14$ \hfill&
\hfill $1.2$ \hfill&
\cr

\+
\hfill C1 \hfill&
\hfill $1.5$ \hfill&
\hfill $0.3$ \hfill&
\hfill $1.75$ \hfill&
\hfill $1.8$ \hfill&
\cr

\+
\hfill C2 \hfill&
\hfill $1.5$ \hfill&
\hfill $0.6$ \hfill&
\hfill $1.63$ \hfill&
\hfill $1.8$ \hfill&
\cr

\+
\hfill C3 \hfill&
\hfill $1.5$ \hfill&
\hfill $0.05$ \hfill&
\hfill $1.79$ \hfill&
\hfill $1.8$ \hfill&
\cr

} 
}
\endtable

After each merger another satellite is generated in the same way, and
then merged.  We measure the profile of the halo by starting at the
potential centre and binning outwards in radius.  The 10\%, 20\%,...
radii are determined at the same time.  Figure~\rquantsfig{}
shows the evolution of the ratios of these decile radii to the
half-mass radius $r_{1/2}$.  This shows that the structure
converges after 7 or 8 mergers.
Figure~\nfitfig{} shows the average profile of the last 7 mergers for
models A, B and C1.  Also shown is the initial Plummer sphere, and a
sanity check: the profile of the initial model, evolved in isolation
for the same time as all the mergers added together in simulation A.
When evolved in isolation the profile changes a
little due to numerical effects, but much less than when mergers are
included.  The cusp slope $\alpha$ was measured in these models by
fitting a power law to the profile at $r<0.3r_{1/2}$.  The results are
given in Table~2 and appear as solid lines in Figure~\nfitfig{}.  The
general prediction that larger $\nu$ produces larger $\alpha$ is
supported, and the actual values of the measured slopes are quite close
to the prediction of equation~(\aleq).

A comparison of the three models with $\nu=1.5$ (C1, C2, C3) shows
that the profiles do not differ significantly.  Thus the final
profile is not sensitive to the value of $\kappa$.

Cosmological halos are not isolated, as our simulated halos are, and
they are not in equilibrium in their outer parts.  Thus we cannot make
direct inferences about the outer profiles of cosmological halos on
the basis of our simulations.  Isolated merger products are expected
to have an outer profile with $\beta=4$ (Jaffe 1987, Merritt, Tremaine
\& Johnstone 1989). Thus in
Figure~\nfitfig{} we see a slow roll-over {}from the
inner cusp to an outer envelope with $\beta\simeq4$, significantly less
steep than the initial Plummer model.  In cosmological halos,
$\beta\simeq3$ is observed near the virial radius for a wide range of
models.  It is not easy to identify the equivalent of the virial
radius in our simulations, but the effective $\beta$ of our profiles
is about 3 in the range $(0.5,2)r_{1/2}$.  The 80\% radius is at about
$2r_{1/2}$, and the 90\% radius (the apocentre of the satellite
orbits) is at about $2.5r_{1/2}$. For comparison, in cosmological
infall models the turnround radius is at about three times the virial 
radius.

We also test the `sinking satellite' assumption of Section~2, which
was used in the calculations of Section~3.1.  If dense satellites
really sink to the centre of the parent then the particles which
originated in the satellite should be concentrated towards the centre
of the merger product.  We test this by measuring the half-mass radius
of the merger product and comparing it with that of the disrupted
satellite particles.  The result is shown in Figure~\rhsfig{}.  The
degree of concentration of the disrupted satellite particles varies as
the profile changes.  The effect is most pronounced when the halos
have a homogeneous core, and the satellite is much denser than the
parent in its inner parts.  As the cusp develops the satellite is
disrupted at a progressively larger radius, as we argued should
happen in Section~2, and the concentration settles down at a value of
around 2 for all the simulations.

\section{Discussion}
The analytic argument leading to equation (\aleq) is highly idealised
in its use of a single power law for the cusp profile.  The
semi-analytic model we have used for merging is based on the same idea
but does not assume a form for the profile.  Nevertheless it is
probably not very accurate in detail.  The $N$-body experiment gives
us confidence that our simple arguments might be close to the
truth.  The values of $\alpha$, measured {}from our experiments, agree
quite well with the prediction of equation (\aleq).  Generally the
semi-analytic experiments give $\alpha$ values higher than predicted,
and the $N$-body experiments give lower values.  

The physical origin we suggest for the cuspy profiles of halos
is that sufficiently dense satellites can sink intact to the halo
centre.  The details of our merger prescription may affect our
detailed conclusions, such as equation (\aleq), but the robustness of
the main physical idea should lead to similar behaviour in
more realistic models.  Figure~\rhsfig{} shows that the basic picture
of Section~2 is not far wrong, at least for the kind of mergers
studied here.

The extent to which this basic picture applies to actual cosmological
halos remains to be demonstrated.  Navarro, Frenk and White (1996)
introduced the NFW profile for CDM halos.  Navarro, Frenk and White
(1997), in agreement with the lower resolution results of Cole and
Lacey (1996), find that the NFW density profile continues to be a good
fit to halos from power-law initial fluctuation spectra.  The expected
trend in the mass-density relation of halos is clearly visible in
their fits: smaller effective $\nu$ for more negative $n$.  According
to the arguments of the present work, we would expect to see a
corresponding trend in profile properties between $n=0$ and $n=-1.5$.
Nevertheless Navarro, Frenk and White (1997) are able to use a single
fitting formula for all their halos.

The apparent contradiction may be accounted for in a number of ways.
Cosmological halos always show a degree of scatter in their measured
properties because of substructure---which is absent by design in our
experiments.  More data on the profiles of similar mass halos would
give better statistical confidence in statements about trends in halo
properties.  in addition, it is difficult to find a consistent and
dynamically meaningful scale for $\rho$ and $r$ on which to base a
comparison of halos.  In Figure~\nfitfig{} the trend between the
profile and $\nu$ is obvious because all the halos line up so well in
their outer parts.  This in turn is a consequence of the fact that
they are dynamically isolated, and hence the half-mass radius and
density can be unambiguously defined.  Since cosmological halos are
not isolated, the same comparison cannot be made in this case.  Next,
the mass-density relation of halos in Navarro, Frenk and White (1997)
has some scatter in it.  If this is a consequence of a real scatter in
the underlying relation it should produce a scatter in halo profile
properties in addition to the effects of substructure.  Finally, our
experiments do not properly account for the expected systematic
dependence on the power spectrum of the distributions of satellite
mass and orbital parameters.  Perhaps these could compensate in some
way for the difference in core structure produced by the different
mass-density relation.  Since in our experiments we find no dependence
of the profile on mass-distribution or orbital eccentricity, and since
to exactly counter the effects of the mass-density relation would
require fine tuning, we do not find this explanation compelling.

One thing is clear {}from our experiments though: it takes only a few
mergers to establish the cusp profile.  This means that the cusp can
adjust to local conditions quite quickly and so should not depend in a
complicated way on the {\em full} merger history of a halo.

\bye